\RequirePackage{fix-cm}
\documentclass[twocolumn,epjc3]{svjour3}  
\smartqed  
\RequirePackage{graphicx}
\RequirePackage[colorlinks,citecolor=blue,urlcolor=blue,linkcolor=blue]{hyperref}
\usepackage{siunitx}
\usepackage{amsmath}
\journalname{Eur. Phys. J. C}
\begin{document}

\title{Reconstructing the neutrino energy for in-ice radio detectors
}
\subtitle{A study for the Radio Neutrino Observatory Greenland (RNO-G)}


\author{
        J. A. Aguilar\thanksref{ULB}
        \and
        P.~Allison\thanksref{OSU}
        \and
        J.~J.~Beatty\thanksref{OSU}
        \and
        H.~Bernhoff\thanksref{UppE}
        \and
        D.~Besson\thanksref{KU, MEPhI}
        \and
        N.~Bingefors\thanksref{UppPh}
        \and
        O.~Botner\thanksref{UppPh}
        \and
        S.~Bouma\thanksref{ECAP}
        \and
        S.~Buitink\thanksref{VUBastro}
        \and
        K.~Carter\thanksref{CalPoly}
        \and
        M.~Cataldo\thanksref{ECAP}
        \and
        B.~A.~Clark\thanksref{MSU}
        \and
        Z.~Curtis-Ginsberg\thanksref{UC}
        \and
        A.~Connolly\thanksref{OSU}
        \and
        P.~Dasgupta\thanksref{ULB}
        \and
        S.~de Kockere\thanksref{VUBelem}
        \and
        K.~D.~de Vries\thanksref{VUBelem}
        \and
        C.~Deaconu\thanksref{UC}
        \and
        M.~A.~DuVernois\thanksref{WIPAC}
        \and
        C.~Glaser\thanksref{UppPh}
        \and
        A.~Hallgren\thanksref{UppPh}
        \and
        S.~Hallmann\thanksref{DESY}
        \and
        J.~C.~Hanson\thanksref{Whittier}
        \and
        B.~Hendricks\thanksref{PSU}
        \and
        B.~Hokanson-Fasig\thanksref{WIPAC}
        \and
        C.~Hornhuber\thanksref{KU}
        \and
        K.~Hughes\thanksref{UC}
        \and
        A.~Karle\thanksref{WIPAC}
        \and
        J.~L.~Kelley\thanksref{WIPAC}
        \and
        S.~R.~Klein\thanksref{LBL}
        \and
        R.~Krebs\thanksref{PSU}
        \and
        R.~Lahmann\thanksref{ECAP}
        \and
        U.~Latif\thanksref{VUBelem}
        \and
        M.~Magnuson\thanksref{KU}
        \and
        T.~Meures\thanksref{WIPAC}
        \and
        Z.~S.~Meyers\thanksref{DESY, ECAP}
        \and
        K.~Mulrey\thanksref{VUBastro}
        \and
        A.~Nelles\thanksref{DESY, ECAP}
        \and
        A.~Novikov\thanksref{KU}
        \and
        E.~Oberla\thanksref{UC}
        \and
        B.~Oeyen\thanksref{Ghent}
        \and
        H.~Pandya\thanksref{VUBastro}
        \and
        I.~Plaisier\thanksref{ECAP, DESY}
        \and
        L.~Pyras\thanksref{DESY, ECAP}
        \and
        D.~Ryckbosch\thanksref{Ghent}
        \and
        O.~Scholten\thanksref{VUBelem, UG}
        \and
        D.~Seckel\thanksref{UD}
        \and
        D.~Smith\thanksref{UC}
        \and
        D.~Southall\thanksref{UC}
        \and
        J.~Torres\thanksref{OSU}
        \and
        S. Toscano\thanksref{ULB}
        \and
        D.~Tosi\thanksref{WIPAC}
        \and
        D.~J.~Van Den Broeck\thanksref{VUBelem, VUBastro}
        \and
        N.~van Eijndhoven\thanksref{VUBelem}
        \and
        A.~G.~Vieregg\thanksref{UC}
        \and
        C. Welling\thanksref{e1,ECAP,DESY}
        \and    
        S.~Wissel\thanksref{PSU, CalPoly}
        \and
        R.~Young\thanksref{KU}
        \and
        A.~Zink\thanksref{ECAP}
}

\thankstext{e1}{e-mail: christoph.welling@desy.de, authors@rno-g.org}


\institute{
            Universit\'e Libre de Bruxelles, Science Faculty CP230, B-1050 Brussels, Belgium \label{ULB}
            \and
            \label{OSU} Dept. of Physics, Center for Cosmology and AstroParticle Physics, Ohio State University, Columbus, OH 43210, USA 
            \and
            \label{UppE} Uppsala University, Dept.~of Engineering Sciences, Division of Electricity, Uppsala, SE-752 37, Sweden
            \and
            \label{KU} University of Kansas, Dept. of Physics and Astronomy, Lawrence, KS 66045, USA
            \and
            \label{MEPhI} National Nuclear Research University MEPhI, Kashirskoe Shosse 31, 115409, Moscow, Russia
            \and
            \label{UppPh} Uppsala University, Dept.~of Physics and Astronomy, Uppsala, SE-752 37, Sweden
            \and
            \label{VUBastro} Vrije Universiteit Brussel, Astrophysical Institute, Pleinlaan 2, 1050 Brussels, Belgium
            \and
            \label{CalPoly} Physics Dept. California Polytechnic State University, San Luis Obispo CA 93407, USA
            \and
            \label{MSU} Dept.~of Physics and Astronomy, Michigan State University, East Lansing MI 48824, USA
            \and
            \label{VUBelem} Vrije Universiteit Brussel, Dienst ELEM, B-1050 Brussels, Belgium
            \and
            \label{UC} Dept.~of Physics, Enrico Fermi Inst., Kavli Inst.~for Cosmological Physics, University of Chicago, Chicago, IL 60637, USA
            \and
            \label{WIPAC} Wisconsin IceCube Particle Astrophysics Center (WIPAC) and Dept.~of Physics, University of Wisconsin-Madison, Madison, WI 53703,  USA
            \and
            \label{ECAP} Erlangen Center for Astroparticle Physics (ECAP), Friedrich-Alexander-University Erlangen-Nuremberg, 91058 Erlangen, Germany
            \and
            \label{DESY} DESY, Platanenallee 6, 15738 Zeuthen, Germany
            \and
            \label{Whittier} Whittier College, Whittier, CA 90602, USA
            \and
            \label{LBL} Lawrence Berkeley National Laboratory, Berkeley, CA 94720, USA
            \and
            \label{PSU} Dept.~of Physics, Dept.~of Astronomy \& Astrophysics, Penn State University, University Park, PA 16801, USA
            \and
            \label{Ghent} Ghent University, Dept. of Physics and Astronomy, B-9000 Gent, Belgium
            \and
            \label{UD} Dept.~of Physics and Astronomy, University of Delaware, Newark, DE 19716, USA
            \and
            \label{UG} KVI – Center for Advanced Radiation Technology, University of Groningen, Groningen, The Netherlands
}

\date{Received: date / Accepted: date}

\maketitle

\begin{abstract}
Since summer 2021, the Radio Neutrino Observatory in Greenland (RNO-G) is searching for astrophysical neutrinos at energies \SI{>10}{PeV} by detecting the radio emission from particle showers in the ice around Summit Station, Greenland. We present an extensive simulation study that shows how {RNO-G} will be able to measure the energy of such particle cascades, which will in turn be used to estimate the energy of the incoming neutrino that caused them. The location of the neutrino interaction is determined using the differences in arrival times between channels and the electric field of the radio signal is reconstructed using a novel approach based on Information Field Theory. Based on these properties, the shower energy can be estimated. We show that this method can achieve an uncertainty of 13\% on the logarithm of the shower energy after modest quality cuts and estimate how this can constrain the energy of the neutrino. The method presented in this paper is applicable to all similar radio neutrino detectors, such as the proposed radio array of IceCube-Gen2.
\end{abstract}

\section{Neutrino detection with radio antennas}
Neutrinos are unique messengers for high-energy astrophysical phenomena. Without an electric charge, they are not deflected by magnetic fields on their way to Earth and as they only interact via the weak force at very low cross-sections, they can travel vast distances undisturbed. Unfortunately, this also means they only have a small chance to interact within any detector one may build to observe them. For the high-energy neutrino observatories, the solution to this problem is to instrument large volumes of water \cite{Adri_n_Mart_nez_2016,Avrorin:2018ijk} or ice \cite{AMANDA:2004eoa,IceCube:2010hwb} and detect Cherenkov light from charged particles created from neutrino interactions. This approach has led to the discovery of an astrophysical neutrino flux \cite{Aartsen:2013jdh} and even the identification of some likely source candidates \cite{IceCube:2018dnn,Stein_2021} for neutrinos beyond the \SI{100}{TeV} energy scale. To explore the energy range beyond that, even larger detection volumes than the \SI{\sim 1}{km^3} of the largest current optical detectors are required. Extending them to the necessary sizes would require large expenses in money and labor, because of the optical properties of water and ice. Deep sea water has an absorption length on the order of tens of meters \cite{Aguilar:2004nw}. While absorption lengths are larger in deep glacial ice, light gets scattered, with scattering lengths also on the order of tens of meters \cite{Ackermann:2006pva}. This limits the maximum possible spacing for optical detection modules, and thereby the affordable size of the detector.

Fortunately, high-energy particle showers in cold ice emit radio signals that are hardly scattered and show attenuation lengths around \SI{\sim 1}{km}\cite{Avva:2014ena,hanson_etal,barwick_besson_gorham_saltzberg_2005}. This allows for the construction of a detector like the Radio Neutrino Observatory in Greenland \cite{Aguilar:2020xnc}, which will monitor a volume of around 100 cubic kilometers, making it a promising candidate for the first detection of EeV neutrinos.

Ultra-high energy neutrinos are expected to be produced by astrophysical sources (e.g.\ \cite{Fang:2013vla,Boncioli_2019,Murase:2014foa,Rodrigues:2020pli,Murase:2007yt}) and by interaction of ultra-high energy cosmic rays interacting with cosmic photon backgrounds \cite{Greisen:1966jv,Beresinsky:1969qj,Zatsepin:1966jv,vanVliet:2019nse}. Cosmogenic neutrino fluxes are predicted to show different energy spectra than neutrinos from astrophysical sources, so distinguishing between them with any radio neutrino detector will require the ability to reconstruct the neutrino energy. The different models for astrophysical sources differ in their predicted high-energy neutrino spectra, so a spectrum measurement would provide valuable information to constrain these scenarios, especially when combined with information from other cosmic messengers. A good energy estimation is also important to efficiently suppress a background of high-energy muons that drops quickly with energy \cite{Garcia-Fernandez:2020dhb}.  

\subsection{Radio emission from particle showers in ice}
When a high-energy neutrino interacts, a fraction of its energy is transferred through deep-inelasting scattering with nucleons, which causes a hadronic shower to develop. In this hadronic shower, among other particles, pions are created. The neutral pions decay into two photons that cause electromagnetic sub-showers. Although the end product that produces the radio emission are many electromagnetic sub showers, these showers are referred to as hadronic showers. If an electron neutrino undergoes a charged current interaction, next to the hadronic shower, an electron is created which initiates an electromagnetic cascade. In that case, two particle showers are present, whose radio signals overlap. Charged current interaction of $\nu_\mu$ and $\nu_\tau$ can occur as well and create $\mu$ or $\tau$ particles, but these do not emit radio signals and are practically invisible for a radio detector by themselves. They may, however, cause secondary particle showers as they decay or propagate through stochastic energy losses which provides a detectable signature \cite{Garcia-Fernandez:2020dhb}. Therefore, in practical terms, the event reconstruction for a radio-based neutrino detector can be thought of as the reconstruction of a particle cascade.

As a particle shower develops in the ice, electrons in the ice are kicked out of their atomic shells and swept along in the shower. Additionally positrons in the shower front annihilate with electrons in the ice.  Both effects cause an excess of negative charges to develop in the shower, which emits radio signals as they propagate \cite{askaryan,jelley}. This process has been confirmed experimentally in several dense media, including ice, at particle accelerators \cite{Saltzberg_2001,Gorham:2004ny,Gorham:2006fy} and in air showers \cite{Schellart:2014oaa,Aab:2014esa}, where it is a subdominant emission process. If an observer is located on the Cherenkov cone (\SI{\sim 56}{^\circ} opening angle in ice), the radio emission of the entire shower development interferes constructively, i.e.\ reaches the observer at the same time, and causes a signal strong enough to be detected with radio antennas. Because of the longer wavelength, the radio signal can still be detectable a few degrees off the Cherenkov cone. As an observer moves away from the Cherenkov cone, the shorter wavelengths lose coherence first, leading to an overall weakening of the signal as well as a change in the shape of the frequency spectrum \cite{AlvarezMuniz:2000fw,AlvarezMuniz:2010ty}.
In the case of a charged-current interaction by an electron neutrino, two showers are created, whose radio signals interfere, leading to a more complex spectrum of the radio signal. At high energies, the electromagnetic shower is elongated by the LPM effect \cite{Landau:1953um,Migdal:1956tc}, which can increase its length to hundreds of meters, compared to the \SI{\sim 10}{m} typical for a hadronic or lower-energy electromagnetic shower, and cause a more irregular shower development \cite{Stanev:1982au,Gerhardt:2010bj,Konishi:1990ya}. This change also affects the radio signal, as signals from different parts of the shower interfere with each other \cite{AlvarezMuniz:1999xx,Glaser:2019cws}. The result is a reduction of the radio emission and an irregular electric field spectrum with multiple maxima.

\subsection{The Radio Neutrino Observatory Greenland}
\begin{figure}
    \centering
    \includegraphics[width=.45\textwidth]{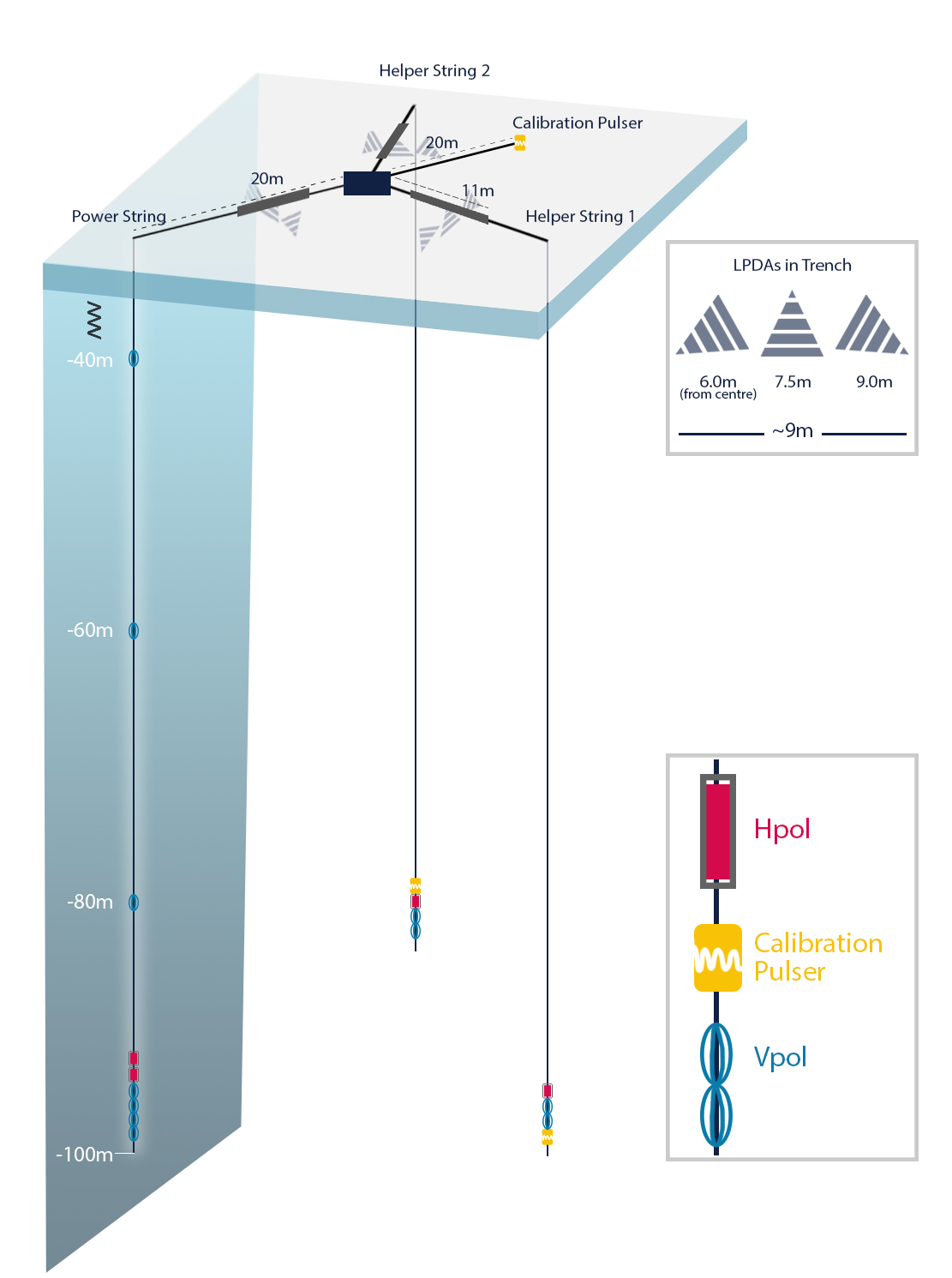}
    \caption{Layout of one of the 35 detector stations that will make up the Radio Neutrino Observatory Greenland (RNO-G). From \cite{Aguilar:2020xnc}}
    \label{fig:RNO_station}
\end{figure}
The Radio Neutrino Observatory Greenland (RNO-G) \cite{Aguilar:2020xnc} is a detector for neutrinos with energies above \SI{10}{PeV}, currently under construction near Summit Station on top of the ice sheet of Greenland. RNO-G is scheduled to be completed with 35 detector stations in 2023, making it the first discovery-scale detector for neutrinos at these energies.

The stations are positioned on a square grid with a spacing of \SI{1.25}{km} between stations. With this spacing, many neutrino events will only be detected by a single station, which maximizes the effective volume of the detector, but presents a challenge for the reconstruction. In this paper, we will assume that only data from a single station is available, effectively turning each station into an independent detector.

An RNO-G station (see Fig.~\ref{fig:RNO_station}) can be seen as consisting of two sections: A \emph{shallow component} with antennas buried about \SI{2}{m} below the snow surface and a \emph{deep component} with antennas placed in holes at a depth of up to \SI{100}{m}. Both are centered around a DAQ box housing the station electronics and a small tower for communication antennas and solar panels.

The shallow component consists of three sets of three logarithmic-periodic dipole antennas (LPDAs) each, installed \SI{2}{m} below the ice surface at a horizontal distance of \SI{11}{m} from the station center. One LPDA in each set is pointed upwards, to detect air showers, while the others are pointed downwards at a \SI{60}{^\circ} angle, making them sensitive to neutrino signals from below. The LPDAs have the advantage of being very sensitive, but they can only be deployed close to the surface because of their size (\SI{140}{cm} height, \SI{146}{cm} width). At these shallow depths, the observable ice volume is smaller because of shielding effects from the changing index of refraction in the upper \SI{100}{m} of the ice, so that only about 20\% of triggered neutrino events are expected to be visible in the shallow component \cite{Aguilar:2020xnc}.

The deep component consists of three vertical cables, called \emph{strings}, in boreholes going down to a depth of \SI{100}{m} below the snow surface. One of the strings, called \emph{power string}, holds a phased array consisting of four vertically polarized (Vpol) antennas. A phased array works by overlaying the signals from all four channels to reduce the noise, which was tested on the ARA detector \cite{Allison:2018ynt} and will allow RNO-G to trigger on radio signals with a signal-to-noise ratio (SNR)\footnote{We define the SNR as half the peak-to-peak voltage amplitude divided by the root mean square of the noise.} as low as 2. Directly above the phased array are two horizontally polarized (Hpol) antennas. Further up the string are three more Vpol antennas with a spacing of \SI{20}{m} from each other.

The other two strings, called \emph{helper strings}, are more sparsely instrumented, with two Vpol and one Hpol antenna on each string. Additionally, each string holds a radio pulser that can be used to calibrate the detector.

The Vpol antenna uses a \emph{fat dipole} design with a height of \SI{60}{cm} and a diameter of \SI{12.7}{cm}, which is sensitive in the \SIrange{\sim 50}{600}{MHz} band and has an antenna response that is symmetric in azimuth. For the horizontal polarization, \emph{quadslot antennas} with a height of \SI{60}{cm} and a diameter of \SI{20.32}{cm} are used, whose design is mostly limited by the diameter of the borehole \cite{Aguilar:2021yQ}. Because of this constraint, the Hpol antennas have a smaller overall sensitivity and are only sensitive in the \SIrange{\sim 200}{400}{MHz} range. Combined with the tendency of the neutrino-induced radio signals to be more vertically polarized and stronger at lower frequencies, this means only a small signal will be visible in the Hpol channels, if at all, for most events.

After a signal is received by one of the downhole antennas, it is fed into a Low Noise Amplifier (LNA) and sent to the surface via a Radio Frequency over Fiber (RFoF) transmitter. At the surface, the signal is amplified again by an LNA inside the DAQ box, digitized and stored on an SD card. This signal chain has a passband of \SIrange{130}{700}{MHz}. From the DAQ the signals are transferred to a server at Summit Station via LTE communication. All components of the signal chain are located in separate RF-tight housings, which helps keep the noise level low, at an expected root mean square (RMS) of \SI{\sim 10}{mV}.

\subsection{General reconstruction strategy}
In general, the amplitude of the electric field $\vec{E}$ of the radio signal from a particle shower is proportional to the shower energy. In practice, it is more convenient to use the energy fluence of the radio signal 
\begin{equation}
    \Phi^E = c \cdot \epsilon \int \vec{E}^2(t) dt,
\end{equation}
where $c$ is the speed of light and $\epsilon$ the permittivity. Its square root is proportional to the shower energy as well, but has the advantage of being easier to determine from data and less affected by fluctuations from noise than the electric field amplitude. 

The other parameters that affect the radio signal received at the detector can be expressed with the relation
\begin{equation}
    \sqrt{\Phi^E} \propto E_\nu \cdot \kappa \cdot \exp\left(-\frac{l}{l_{\text{att}}}\right) / l \cdot f(\varphi)
    \label{eq:energy_reco_formula}
\end{equation}
where $E_\nu$ is the neutrino energy, $\kappa$ is the fraction of the neutrino energy that is transferred into the shower, $\varphi$ is the viewing angle, $l$ is the distance the radio signal travels in the ice before reaching the antenna and $l_{\text{att}}$ is the attenuation length of the ice. Based on this, we will first reconstruct geometric shower parameters and the energy of the particle shower, and then use this to estimate the neutrino energy.

This study focuses on the reconstruction based on the deep component of RNO-G. Due to the phased array trigger, this component is expected to detect most neutrino signals. An energy reconstruction study for the ARIANNA experiment has been performed previously \cite{Anker_2019,Glaser:2019kjh}. Since the surface component of RNO-G has been modeled after the ARIANNA experiment, the reconstruction strategy should be similarly applicable for potential events detected with the LPDAs only.

\section{Influence of the viewing angle}
\label{sec:viewing_angle}
\begin{figure}
    \centering
    \includegraphics[width=.45\textwidth]{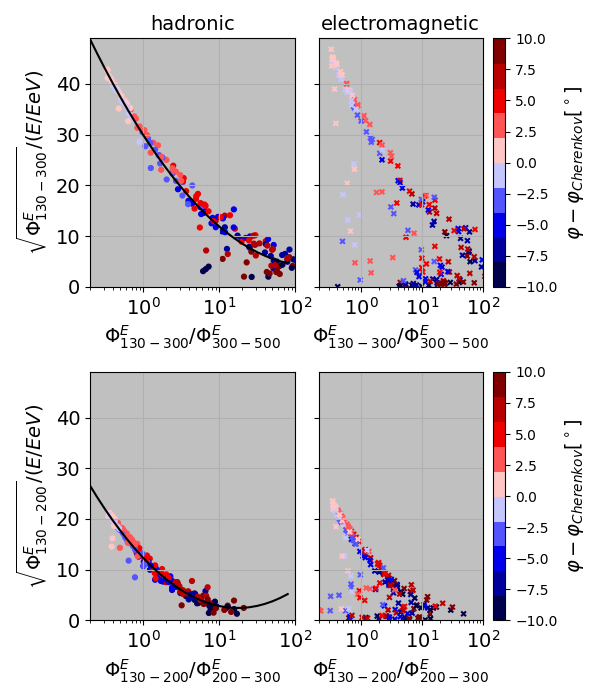}
    \caption{Square root of the energy fluence of the radio signal emitted by a particle shower in ice divided by the shower energy as a function of the ratio between the energy fluence in different passbands for hadronic (left) and electromagnetic (right) showers. Radio signals were generated using the ARZ2019 model \cite{PhysRevD.84.103003} for shower energies between \SI{1e15}{eV} and \SI{1e19}{eV}. The upper plot uses a lower passband of \SIrange{130}{300}{MHz} and an upper passband of \SIrange{300}{500}{MHz}, the lower plot uses \SIrange{130}{200}{MHz} and \SIrange{200}{300}{MHz}. Colors show the viewing angle relative to the Cherenkov angle. The viewing angle is calculated with respect to the position of the maximum of the charge-excess profile. The black line shows the parametrization for the hadronic showers.}
    \label{fig:viewng_angle_parametrization}
\end{figure}

The last term in Eq.~\ref{eq:energy_reco_formula} parametrizes the contribution of the \emph{viewing angle}, defined as the angle between the direction into which the signal is emitted and the shower axis.
At the Cherenkov angle, the radio emission over the entire shower development interferes constructively, leading to a strong signal. As the viewing angle moves away from the Cherenkov angle, the radio signal gets weaker, but not equally over the whole frequency range. Shorter wavelengths lose coherence faster, which allows us to use the shape of the frequency spectrum of the radio pulse as a proxy for the viewing angle. This technique has already been demonstrated to work for air showers \cite{Welling:2019scz}, so we adapt it here for neutrino detection.

To quantify the shape of the frequency spectrum, we define a \emph{slope parameter}
\begin{equation}
    \label{shape_parameter_definition}
    s = \Phi^E_{130-300} / \Phi^E_{300-500}
\end{equation}
as the ratio between the energy fluence of the radio signal in the \SIrange{130}{300}{MHz} and the \SIrange{300}{500}{MHz} passbands.  More details of the procedure to reconstruct the electric field and the parameter $s$ are discussed in Sec.~\ref{sec:efield}.

The relation between the slope parameter and the energy fluence of the radio signal is shown in the upper plot of Fig.~\ref{fig:viewng_angle_parametrization} for a fixed distance between shower and observer of \SI{1}{km} and neglecting any attenuation by the ice. It is convenient to only use the energy fluence in the lower passband to estimate the shower energy, because it is less dependent on the viewing angle than higher frequencies, and also easier to reconstruct, because of the higher sensitivity of the Vpol antennas at lower frequencies.

For hadronic showers, the relation between $\lg(s)$ and $\sqrt{\Phi^E} / E$ can be parameterized by a parabola, for electromagnetic showers the relation is more complicated: The radio emission stems from the electrons and positrons in the shower, so the signal from  electromagnetic showers tends to be a bit stronger than for hadronic showers of the same energy. However, outliers with a much weaker signal originate from showers affected by the LPM effect. These showers are elongated and/or show an irregular development, which can effectively lead to several shower maxima. This causes a loss of coherence of the radio signal emitted as the shower develops, thus, weakening it. It should also be noted that for such long showers, an observer will see different parts at different viewing angles, making this parameter rather ill-defined.

Since no method to distinguish between neutrino flavors has yet been developed for RNO-G, we assume that no information about the flavor or the interaction type (charged vs. neutral current) is available. We therefore only parameterize the relationship the between $\sqrt{\Phi^E}/E$ and $s$ for hadronic showers with a parabola, and then use the result for both shower types.

If the difference between viewing and Cherenkov angle is large, the electric field spectrum can drop to close to 0 in the \SIrange{300}{500}{MHz} band. The parametrization still holds for these cases, but even small uncertainties in the shape of the reconstructed spectrum can lead to large changes in the $s$ parameter. We found empirically that we can mitigate this problem by using a  slope parameter $s^\prime$ if $s$ is larger than 10. This new slope parameter $s^\prime$ is defined similarly to $s$, but using the energy fluences in the \SIrange{130}{200}{MHz} and the \SIrange{200}{300}{MHz} bands, which can be parameterized the same way as $s$ (see Fig.~\ref{fig:viewng_angle_parametrization}, bottom). The values of the parameters of the parabola fit are listed in Tab.~\ref{tab:spectrum_fit_parameters}.

Using this parametrization, the shower energy can be calculated from the radio signal using
\begin{equation}
\begin{aligned}
    E_{\text{sh}} = \frac{\sqrt{\Phi^E_{130-300} / \text{eV}} }{p_2 \cdot \lg(s)^2 + p_1 \cdot \lg(s) + p_0} \cdot \text{EeV} \\
     = \sqrt{\Phi^E_{130-300}/\text{eV}} \cdot f_\varphi (s) \cdot EeV
\end{aligned}
    \label{eq:spectrum_shape_formula}
\end{equation}
where $p_i$ are the parameters of the parabola from Tab.~\ref{tab:spectrum_fit_parameters}. If the $s$ parameter is larger than 10, $\Phi^E_{130-200}$ and $s^\prime$ are used instead. The points in Fig.~\ref{fig:viewng_angle_parametrization} are not exactly on the parametrization line. We can calculate the shower energy from Eq.~\ref{eq:spectrum_shape_formula} for these showers using the true values of $ \Phi^E$ and s and compare them to the true shower energy. The difference is an estimate of the relative uncertainty on the shower energy we get from the chosen parametrization. Using the 68\% quantile as a criterion, the relative uncertainty is 9\%, which will turn out to be a subdominant uncertainty.
However, these calculations were done at a fixed distance of \SI{1}{km} and without the effects of attenuation. So before this equation can be used, the electric field needs to be reconstructed and the distance to the shower needs to be known to correct for attenuation during propagation.

\begin{table}
    \centering
    \begin{tabular}{|c|c|c|c|}
    \hline
    & $p_2$ & $p_1$ & $p_0$\\
    \hline
    $s$ & $4.70$ & $-22.20$ & $29.99$ \\
    $s^\prime$ & $6.46$ & $-16.00$ & $12.49$ \\
    \hline
    \end{tabular}
    \caption{Parameters of the parabola parametrization to the hadronic showers shown in Fig.~\ref{fig:viewng_angle_parametrization} that are used to reconstruct the shower energy from Eq.~\ref{eq:spectrum_shape_formula}.}
    \label{tab:spectrum_fit_parameters}
\end{table}

\section{Vertex reconstruction}
After a neutrino event has been identified, the first step in the reconstruction process is to determine the location of the neutrino interaction. This is important for the energy reconstruction in order to be able to correct for the attenuation stemming from the signal propagation through the ice and the $\frac{1}{r}$ weakening of the signal amplitude with distance. Knowing the vertex location also allows us to determine the angle at which the radio signal arrives at the detector, which determines the antenna response.

\subsection{Radio signal propagation through ice}
The index of refraction $n$ of glacial ice is not uniform but decreases with depth.  Measurement from the South Pole, Moore's Bay and Summit Station \cite{Barwick_2018,Deaconu_2018bkf} all suggest that the index of refraction profile can be parameterized as a function of the depth $z$ by
\begin{equation}
    n(z) = n_0 - \Delta_z e^{z/z_0}
    \label{eq:index_of_refraction}
\end{equation}
where $n_0$, $\Delta_z$ and $z_0$ are fit parameters that have to be determined from measurements. Because of this, the radio signals do not travel in a straight line but are bent downwards.
Additionally, a radio signal may be reflected at the ice surface and reach the detector this way. This lets us divide the ray trajectories into three categories:
\begin{itemize}
    \item \emph{direct ray}: The ray path is bent, but the depth is monotonically in- or decreasing over the entire trajectory.
    \item \emph{refracted ray}: The ray starts going upwards, but eventually bends downwards and reaches the detector from above.
    \item \emph{reflected ray}: The ray is reflected at the ice surface and reaches the detector from above.
\end{itemize}
In many cases, two of these solutions exist and the antenna detects two radio pulses from the same particle shower. These kinds of events, dubbed \emph{DnR} events, become more likely with increasing vertex distance, as the launch angles for both ray paths become more similar, making it more likely for both to be close enough to the Cherenkov angle for a strong enough signal.

Another consequence of the changing index of refraction is the existence of so-called \emph{shadow zones} from where the refraction prevents the radio signal from reaching the antenna at all. These \emph{shadow zones} become smaller the deeper in the ice the antenna is placed, so in many cases only the deeper antennas of the detector are able to receive signals.

Contrary to these naive assumptions, horizontal propagation of radio signals from the shadow zones has been observed at the South Pole, the Ross Ice Shelf as well as Summit Station in Greenland \cite{Barwick_2018,Deaconu_2018bkf}, likely due to local deviations from a smooth density profile. However, the measurements at Summit Station suggest that this propagation mode only has a coupling at the percent level, so it will be ignored in this paper.
Additionally, birefringence may lead to a direction- and polarization-dependent index of refraction, which has been observed at the South Pole \cite{KRAVCHENKO2011755}. As of the time of writing, no such measurements are available for Summit Station, so this effect will be ignored here as well.

Assuming an index of refraction profile of the form as in Eq.~\ref{eq:index_of_refraction}, it is possible to find an analytic solution to the ray tracing problem \cite{Glaser:2019cws}. For more complex profiles, other methods are available \cite{Winchen:2019uar,Prohira:2020lmg}, but at the cost of larger computing times.

\subsection{Reconstruction method}
\begin{figure}
    \centering
    \includegraphics[width=.45\textwidth]{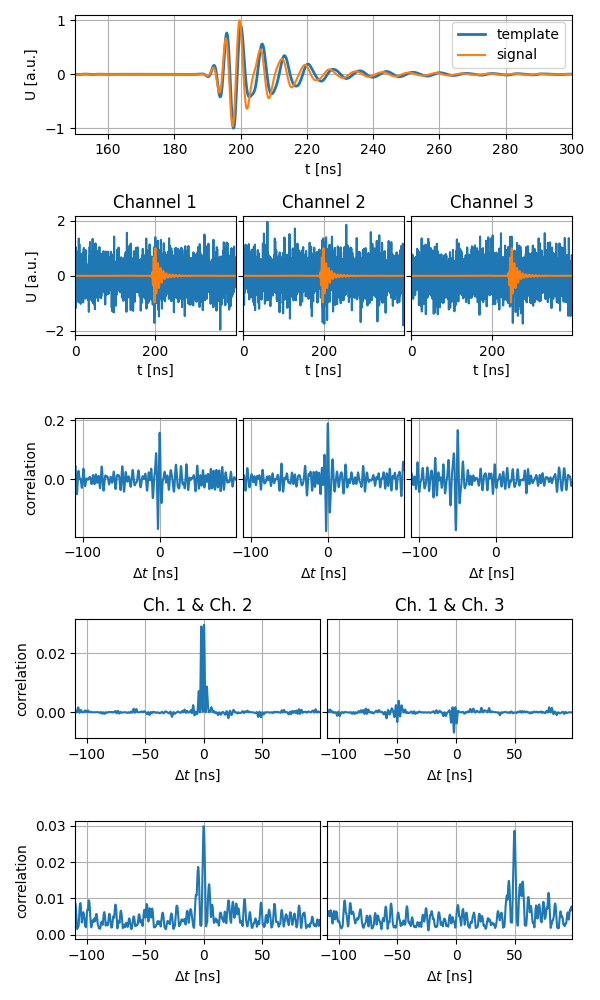}
    \caption{Visualization of the template correlation method to determine the timing differences between channels. Top row: Zoom of waveforms of signal and template. Second row: Noisy waveforms (blue) and the original signals (orange) of three different channels. Third row: Correlation of those waveforms with the template. Fourth row: Product of the template correlation for channel 1 with those for the other two channels. Bottom row: Maximum of the product of the correlations for two channels as a function of the time shift applied to one of them.}
    \label{fig:template_correlation}
\end{figure}
To reconstruct the position of the interaction vertex, we first present a method to determine the relative timing between waveforms in two different channels, then we show how to use this knowledge for the vertex reconstruction.

Most neutrino events detected by RNO-G are expected to be just above the noise threshold, so a method that also works at low signal-to-noise ratios is needed. In the first step we apply a 10th order Butterworth filter with a passband of \SIrange{130}{300}{MHz} to the waveforms to remove frequencies where noise is likely to be dominant over the actual radio signal from the shower.

A convenient way to determine the timing difference between two waveforms is to calculate the correlation function between them, given by the equation
\begin{equation}
    \rho(\Delta n) = \frac{\sum_i (V_1)_i \cdot (V_2)_{i-\Delta n}}{\sqrt{\sum_i (V_1)_i^2}\cdot \sqrt{\sum_i(V_2)_i^2}}
\end{equation}
where $\Delta n$ is the number of samples by which one waveform has been shifted and $V_1$ and $V_2$ are the voltages measured by both channels. The denominator is a normalization, so that $\rho(\Delta n)=1$, if both waveforms are identical.    
However at low SNRs, performance of this method diminishes, as spurious correlations between noise become likely. To avoid this, we use a \emph{template} that was derived by folding an electric field with the detector response, as discussed in the next paragraph. This process is visualized in Fig.~\ref{fig:template_correlation}: The top row shows three waveforms (orange) to which random noise was added, resulting in a noisy signal (blue). The three waveforms are identical, except that channel 3 was shifted by \SI{50}{ns}. The second row shows the correlations of these waveforms with a template, which is identical to the original waveform as well in this example. In the next step (third row) we combine two of these correlations by multiplying them with each other. Note that for channels 1 and 2 there is a sharp peak, which is missing for channels 1 and 3. This is because the timing difference between channels 1 and 2 is 0, so their correlations have their maxima at the same position. Because of the \SI{50}{ns} timing difference between them, the correlation maxima for channels 1 and 3 are at different positions and disappear when both are multiplied. So for there to be a peak in their product, one of the correlations would have to be shifted by \SI{50}{ns}. Turning this reasoning around, we can determine the time difference between two waveforms by calculating their correlation functions with a template and shifting one of the correlations to find when their product has the largest maximum. This is shown in the bottom row of Fig.~\ref{fig:template_correlation}: For channels 1 and 2 the product of their correlation functions (from here on called \emph{correlation product}) has its highest maximum at a time shift of \SI{0}{ns}, for channels 1 and 3 it is at \SI{50}{ns}, which is the correct result. By using the correlations to a template, the height of the maximum is also weighted by how \emph{signal-like} the channel waveforms are, which will become useful later. While RNO-G records data with a sampling rate of \SI{2.4}{GHz}, the Nyquist-Shannon sampling theorem \cite{1697831} allows us to increase the timing precision by resampling the waveform. We use an upsampling to \SI{5}{GHz} to achieve a time binning of \SI{0.2}{ns}. The advantage this approach has compared to correlating the waveforms directly is that it avoids spurious correlations in the noise. Even at SNRs as low as 2.5, the time shift between waveforms can be determined with sub-nanosecond precision.

\begin{figure}
    \centering
    \includegraphics[width=.45\textwidth]{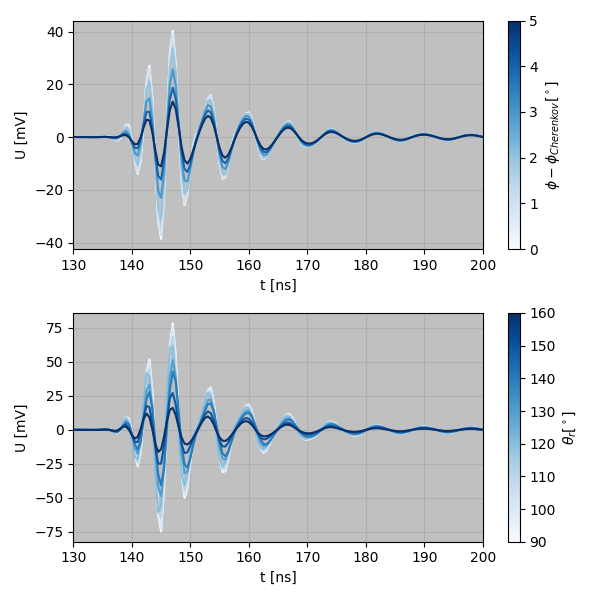}
    \caption{Examples of waveforms measured by an RNO-G Vpol channel, filtered to the \SIrange{130}{300}{MHz} passband, for different viewing angles (top) and zenith angles of the signal arrival direction (bottom).}
    \label{fig:pulse_shapes}
\end{figure}

This leads to the question what template to use. In the previous example, the template was identical to the signal, but in reality, the signal depends on the viewing angle and the direction from which the signal reaches the antenna. Fortunately, in the relatively small band of \SIrange{130}{300}{MHz}, the shapes of the waveforms are mostly determined by the response of the antenna, amplifier and filters which are very well known. This is demonstrated in Fig.~\ref{fig:pulse_shapes}: While changes in the viewing angle and the signal arrival direction change the height of the individual peaks of the waveform, the position of each peak remains unchanged and their correlations with another waveform will have their maxima at the same positions as well. Therefore, the choice of which template to use has only a minor influence on the result of the timing reconstruction.

With a way to identify the relative timing of the radio signals, the next step is to use them to reconstruct the location of the neutrino vertex. Because of refraction, calculating the signal travel times from a given point to one of the antennas is not straightforward, and somewhat computationally expensive. Therefore, for each antenna depth used by RNO-G, a lookup table with travel times for a grid of possible vertex positions is created. By assuming radial symmetry of the propagation times, only the radial distance and the depth of the vertex position relative to the antenna need to be considered\footnote{To date, no measurements from Summit Station suggest any significant radial asymmetry for ray propagation. Should this change, the lookup tables would need to be adapted, which increases their size and computational cost, but is not a fundamental problem.}. Because of the changing index of refraction, two ray tracing solutions exist, so the propagation times for both are stored. Using lookup tables has two advantages: It saves time, because the ray tracing only has to be done once, and it does not make any assumptions about the propagation. Moreover, we are able to calculate the propagation times for any given vertex position. So if changes to the ice model are necessary or more advanced ray tracing methods become available, all that is necessary is to update the lookup tables. For this paper, a grid spacing of \SI{1}{m} in horizontal distance and \SI{2}{m} in depth was used. This provides adequate precision, though we did not test how a larger spacing would impact the results.

With these lookup tables, the difference in signal arrival times between two channels can be determined for any potential vertex position. Then the product of template correlations for the corresponding time shift is calculated. By doing this for all channel pairs, the sum of all correlations can be used as an estimator for how well the signal timings match a given vertex position.

At this point, we do not know if the radio signal reached the detector via the direct, reflected or refracted path. While we know which channels measured a pulse above a certain SNR, a signal may still be present in the others, but hidden in the noise. Adding the result of the correlation product for channels without a signal or assuming the wrong ray type fortunately does not have serious negative effect on the reconstruction. The correlation of the template with pure noise is so small that it has little influence on the overall correlation sum for all channel pairs. If the wrong ray type is used, the correlation product may have a large enough maximum at a specific vertex position, but this point is usually far enough away from where other channel pairs have their maximum to not affect the reconstruction. Furthermore, for the correct ray types, all channel pairs with a signal will have their largest correlations corresponding to the same point, while wrong ray types or spurious correlations will result in large correlations at different points for each channel pair. Therefore, we use all possible pairs of Vpol channels together, and sum up the correlations for all possible combinations of ray types. The only condition is that the two channels are sufficiently far away from each other. If both channels are part of the phased array or on the same helper string, they are too close to be useful for vertex reconstruction, and therefore not used.

\begin{figure}
    \centering
    \includegraphics[width=.45\textwidth]{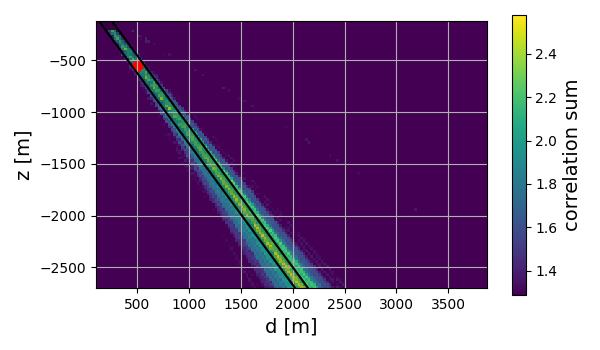}
    \caption{Scan to restrict the vertex search volume. Results are projected in the horizontal distance v. depth plane, with the colors showing the maximum correlation sum over all azimuths. The black lines mark the volume in which the finer scan will be performed. The red dot marks the true vertex position.}
    \label{fig:2d_correlation_map}
\end{figure}

In principle, we could now do a scan on a 3D grid over the entire detector volume, sum up the correlation products for all channel pairs at each point and find its maximum, but this would take too long and consume too much memory to be practical. Instead, we first perform a scan on a grid with a larger spacing. A convenient choice is to use cylindrical coordinates with a spacing in radius of \SI{100}{m}, \SI{2.5}{^\circ} in azimuth and \SI{25}{m} in depth, and calculate the correlation sums for all points on this grid. We flatten this grid into 2 dimensions by taking the maximum correlation sum over all azimuths for each pair of radii and depths and fit a line to the radius vs. depth graph of the points with the largest correlation sums. To identify the azimuth of the neutrino vertex position, we sum up the correlations for all points belonging to the same azimuth and find the maximum. The result is a line in 3D space around which be lay a cuboid that marks the search volume (see Fig.~\ref{fig:2d_correlation_map}).

\begin{figure}
    \centering
    \includegraphics[width=.45\textwidth]{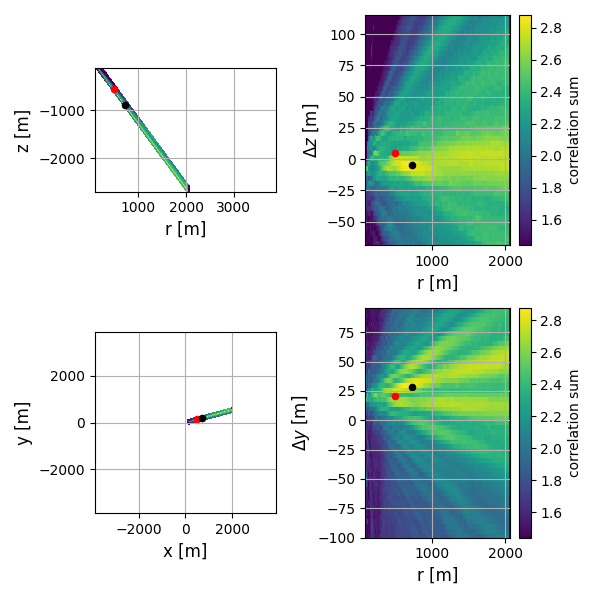}
    \caption{Vertex reconstruction. Results are projected into the horizontal distance vs. depth (top) or east vs. west plane (bottom), with the colors showing the maximum correlation sums over all azimuths and depths, respectively. The left side shows results in coordinates relative to the station position. Zones that are outside the search volume are in white. The right shows coordinates relative to the vertex search region. The red dots mark the true vertex position, the black dots the position with the largest correlation sum.}
    \label{fig:correlation_map}
\end{figure}

Now the search volume is small enough to perform a scan over a smaller grid within a reasonable computational envelope. The result is shown in Fig.~\ref{fig:correlation_map}. The scan was done in 3D but we again only show the projection onto the horizontal distance $r$.

\begin{figure}
    \centering
    \includegraphics[width=.45\textwidth]{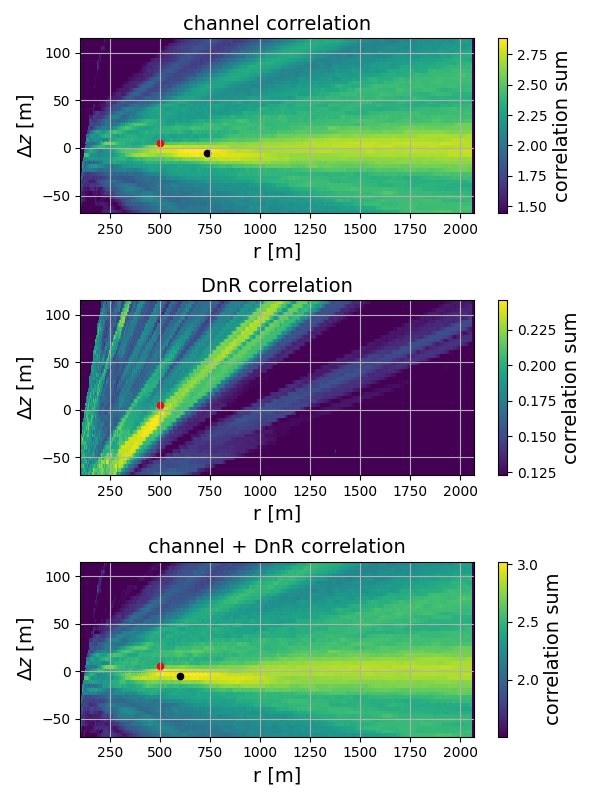}
    \caption{Use of DnR pulses for vertex reconstruction. Top: Correlation sum using only the correlation products between channels. Middle: Correlation sum using DnR pulses. Bottom: Sum of both correlation between channels and DnR pulse correlation. Results are shown projected into the horizontal distance vs. depth plane, with colors showing the maximum correlation sum over all azimuths. The red dot marks the actual vertex position, the black dot the point with the largest correlation sum.}
    \label{fig:dnr_correlation_map}
\end{figure}

This already yields good results, but can be further improved if DnR pulses are present. In most cases with DnR pulses, only one of them will be clearly visible while the other is hidden in the noise. However, even these small signals can still yield valuable information. Using the time difference between DnR pulses works the same way as before, except this time the template correlation is done for a channel with itself. Like for the correlation between channels, we use the correlation products from all channels, as they are already weighted by the correlation with a template and spurious correlations usually disagree with the results from other channels, so they do not pose a problem. The effect of this can be seen in Fig.~\ref{fig:dnr_correlation_map}. Similarly to the correlations between channels, DnR correlations also have the shape of lines, but there are two advantages that make them very useful when combined with the results from correlating different channels: The time differences between DnR pulses change much more with vertex distance than the differences between channels, leading to much thinner lines. The lines are also at a different angle, so the point where they cross with the lines from the correlation sums between channels can be very well constrained. The contribution of the DnR pulses to the correlation sum is much smaller than from the channel pairs, for two reasons: The two pulses tend to differ in amplitude, so one of them is often just above, or even hidden under, the noise, so their correlation is small. Additionally the number of channel pairs is larger than the number of channels.

\subsection{Performance of vertex reconstruction}
\label{sec:vertex_reco_quality}
To assess the performance of the neutrino vertex reconstruction (and of the energy reconstruction steps following after this), we produce a simulated data set using the NuRadioMC \cite{Glaser:2019cws} and NuRadioReco \cite{Glaser:2019rxw} software packages. First, we generate a set of $10^7$ neutrinos with isotropic directions, which interact at randomly selected points inside a cylinder with a radius of \SI{3.9}{km} and a depth of \SI{2.7}{km} below the ice surface (corresponding roughly to the ice thickness at Summit Station). Energies are randomly distributed between \SI{5.e16}{eV} and \SI{1.e19}{eV} with a spectrum following the extension of the spectral index measured by IceCube \cite{Haack:2017E1} combined with a model of the expected flux from the interaction of ultra-high energy cosmic rays with the cosmic microwave background and other photon fields \cite{vanVliet:2019nse}. Interactions are generated assuming a 1:1:1 flavor mixing due to oscillations and a probability of $0.71$ to interact via charged current interaction. Absorption of the neutrino in the Earth is taken into account by applying a weighting to events when evaluating the results later.

For each generated shower, the radio emission is calculated using the ARZ2019 model \cite{PhysRevD.84.103003} and propagated to each detector antenna using an analytic ray tracing method \cite{Glaser:2019cws}. If the radio signal reaches an antenna, it is folded with the response of the antenna and of the amplifier. As a proxy for the phased array trigger of RNO-G, the trigger is simulated as a simple high-low threshold trigger that triggers if the voltage exceeds a maximum of \SI{20}{mV} and a minimum of \SI{-20}{mV} in both of the lower two channels of the phased array. If a trigger signal is given, Rayleigh-distributed noise with an RMS of \SI{10}{mV} is added to the waveform of each channel. This means triggering at the $2\sigma$ level, which is the expected performance of the full phased array trigger. It should be noted that the trigger is performed before noise is added. With the real detector, most triggers will be from noise and have to be removed. But as event identification is not the focus of this paper, we assume a 100\% pure event sample.

Finally, the waveforms are upsampled to a sampling rate of \SI{5}{GHz} and the vertex reconstruction is performed.

\begin{figure}
    \centering
    \includegraphics[width=.45\textwidth]{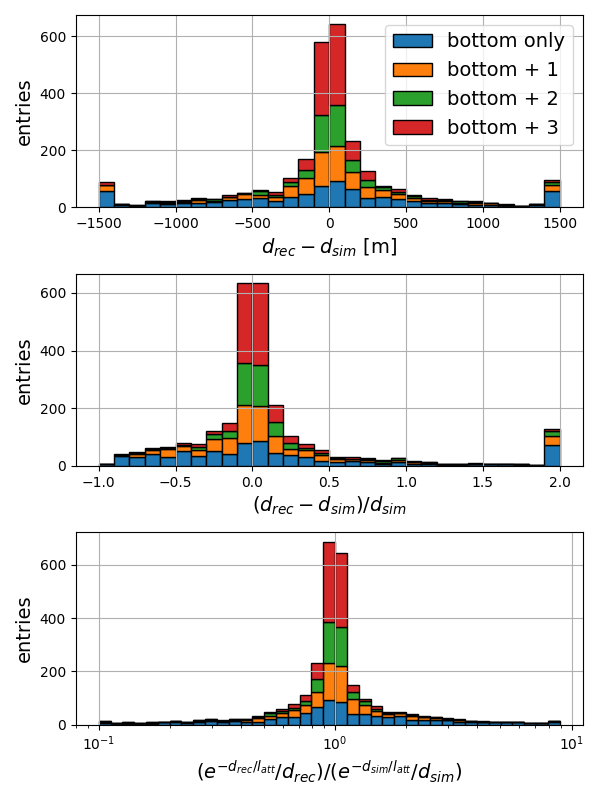}
    \caption{Stacked histograms showing the performance of the vertex reconstruction method. Top: Difference between reconstructed and actual distance between shower and station. Middle: Relative uncertainties on the reconstructed distance. Bottom: Ratio between the expected attenuation of the radio signal using the reconstructed and the actual distance to the shower. Events are divided into four categories: Those where only channels of the phased array or on the helper strings detected a signal with SNR $>2.5$, and those where 1, 2, or 3 of the additional channels on the power string detected a signal with SNR $ > 2.5$. The outermost bins in all plots are overflow bins.}
    \label{fig:distance_error_histograms}
\end{figure}

\begin{table}[]
    \centering
    \begin{tabular}{|c|c|c|c|}
        \hline
         & btm \& 1 & btm \& 2 & btm \& 3  \\
         \hline
         \multicolumn{4}{|l|}{$d_{\text{rec}} - d_{\text{sim}}$} \\
         \hline
         median [m] & 38 & 28 & 0 \\
         $\sigma_{68\%}$ [m] & [-118, 469] & [-91, 235] & [-80, 115] \\
         \hline
         \multicolumn{4}{|l|}{$\Delta d / d_{\text{sim}}$} \\
         \hline
         median & 0.04 & 0.03 & 0.00 \\
         $\sigma_{68\%}$ & [-0.11, 0.59] & [-0.07, 0.29] & [-0.06, 0.11] \\
         \hline
         \multicolumn{4}{|l|}{$\text{att}_{\text{rec}} / \text{att}_{\text{sim}}$} \\
         \hline
         median & 0.96 & 0.97 & 1.00 \\
         $\sigma_{68\%}$ & [0.63, 1.12] & [0.77, 1.08] & [0.90, 1.07]\\
         \hline
         \hline
         &bottom & btm \& 1+ & btm \& 2+ \\
         \hline
         \multicolumn{4}{|l|}{$d_{\text{rec}} - d_{\text{sim}}$} \\
         \hline
         median [m] & 7 & 8 & 8 \\
         $\sigma_{68\%}$ [m] & [-302, 287] & [-156, 204] & [-85, 146] \\
         \hline
         \multicolumn{4}{|l|}{$\Delta d / d_{\text{sim}}$} \\
         \hline
         median & 0.01 & 0.01 & 0.01 \\
         $\sigma_{68\%}$ & [-0.25, 0.35] & [-0.12, 0.21] & [-0.06, 0.14] \\
         \hline
         \multicolumn{4}{|l|}{$\text{att}_{\text{rec}} / \text{att}_{\text{sim}}$} \\
         \hline
         median & 0.99 & 0.99 & 0.99 \\
         $\sigma_{68\%}$ & [0.74, 1.33] & [0.83, 1.14] & [0.87, 1.07]\\
         \hline
    \end{tabular}
    \caption{Median and 68\% quantiles of the absolute and relative uncertainties on the vertex distance, and the ratio between reconstructed and actual attenuation factor $\exp(-d/l_{\text{att}})/d$. In the upper table, events are divided into categories where at least one of the channels at \SI{\sim 100}{m} depth has detected a pulse with SNR $\geq2.5$, as well as exactly one, exactly two or exactly three of the upper channels on the power string. In the lower table, categories where at least one of the channels at \SI{\sim 100}{m} depth has detected a signal and where \emph{at least} one, two or three of the channels further up the power string detected a signal with SNR $\geq 2.5$.}
    \label{tab:vertex_reco_resolution}
\end{table}

The results are shown in Fig.~\ref{fig:distance_error_histograms} and Tab.~\ref{tab:vertex_reco_resolution}. The signal arrival direction is much easier to reconstruct than the vertex distance, so we will only focus on the distance here. It turns out that for the performance of the vertex reconstruction, the SNR of the waveforms matters less than which channels detect a signal at all. Therefore it is useful to divide events into categories depending on which channels recorded a signal with $\text{SNR} > 2.5$, which is the SNR around which the signal becomes large enough to be reliably identified using the template correlation method.

\begin{figure}
    \centering
    \includegraphics[width=.45\textwidth]{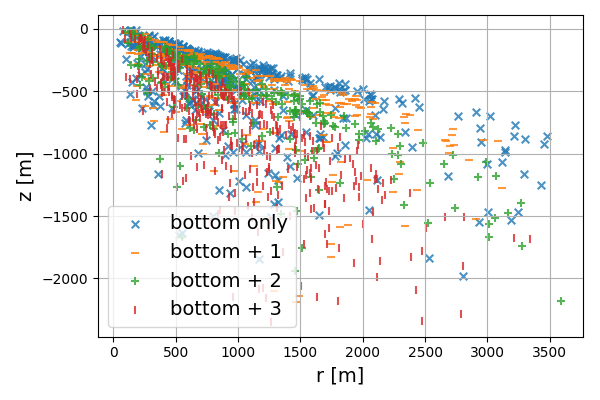}
    \caption{Distribution of shower locations relative to the detector station for events that have signals with SNR $>2.5$ only for the channels at \SI{\sim 100}{m} depth and for 1, 2, or 3 of the other channels on the \emph{power string} as well. For better readability, only every second event is drawn.}
    \label{fig:event_geometries}
\end{figure}

Because the phased array trigger is at the bottom of the \emph{power string} and the interaction vertex can lie in the \emph{shadow zone} of the channels further up for some event geometries, the channels at \SI{\sim 100}{m} depth are the most likely to detect a signal. Because they are all at roughly the same depth, their signal amplitudes, and therefore their SNRs, tend to be similar. As Fig.~\ref{fig:distance_error_histograms} shows, only the channels at the bottom of the detector are usually not enough to reconstruct distance to the vertex. If a signal with $\text{SNR}>2.5$ is available from one more channel (typically the antenna at \SI{80}{m}), the distance reconstruction is already good enough for the uncertainty on the signal attenuation to be better than a factor of 2. Additional channels improve the results even further.

\begin{figure}
    \centering
    \includegraphics[width=.45\textwidth]{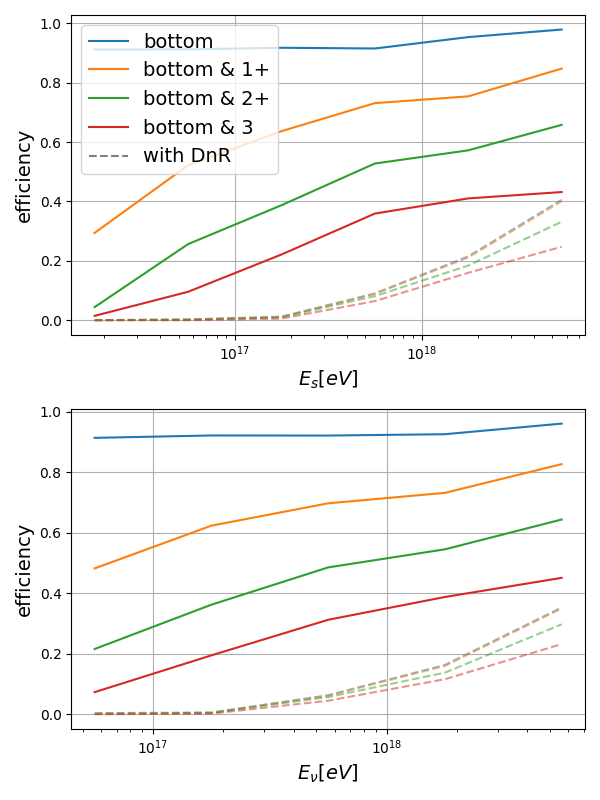}
    \caption{Reconstruction efficiency based on a simplified trigger simulation. Probability for an event to have at least one channel with SNR $> 2.5$ in one of the channels at \SI{\sim 100}{m} depth and to have SNR $>2.5$ in the additional channels on the \emph{power string} as well, as a function of shower energy (top) and neutrino energy (bottom). The dotted lines show the probability that at least one channel detected a second pulse that also has SNR $>2.5$.}
    \label{fig:efficiencies}
\end{figure}

\begin{figure}
    \centering
    \includegraphics[width=.45\textwidth]{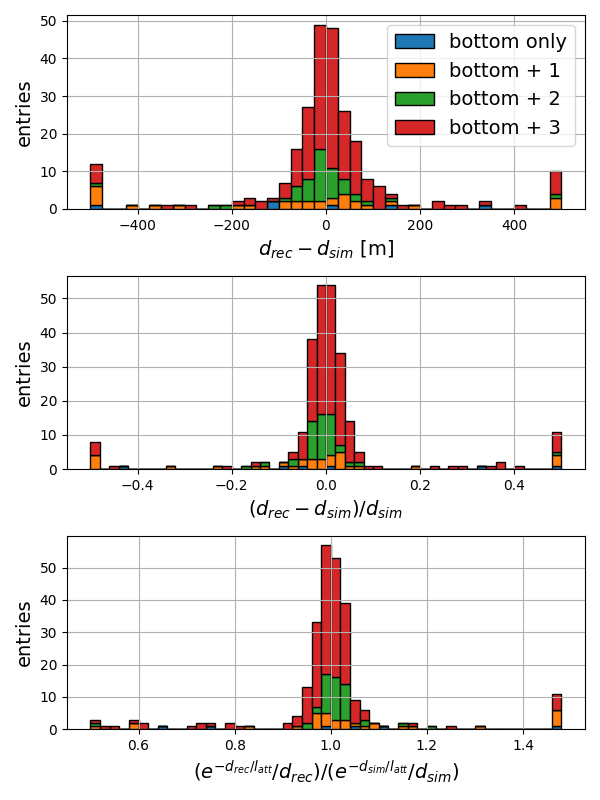}
    \caption{Stacked histograms showing the performance of the vertex reconstruction for events where at least one channel detected two radio signals (DnR) with SNR $>2.5$. Top: Difference between reconstructed and actual distance from station to shower. Middle: Relative uncertainty on the distance reconstruction. Bottom: Ratio of reconstructed and actual attenuation of the radio signal on its way from shower to detector. Events are divided into the same categories as in Fig.~\ref{fig:distance_error_histograms}, but a much smaller subset of events (see Fig.~\ref{fig:efficiencies}) shows DnR signals.}
    \label{fig:dnr_resolution}
\end{figure}

Fig.~\ref{fig:event_geometries} shows the distribution of vertex locations grouped per number of channels with a signal. The figure shows that rather than simply distance, the vertex location most prominently determines which of the upper channels detect a signal, as the vertex may simply be in the shadow zone for the channels further up. The shower energy plays a role as well, because channels at different depths will see the shower at slightly different viewing angles. So at least some of them have to be further off the Cherenkov angle and require a larger shower energy for the radio signal to still be strong enough to be detected. Additionally, showers with lower energies need to be closer to the station to be detectable, translating into larger differences between viewing angles. Fig.~\ref{fig:efficiencies} shows the fraction of events falling into each of these categories as a function of energy relative to all triggered events. It also shows the probability for a DnR pulse to be detected by at least one channel, which greatly improves the vertex reconstruction (see Fig.~\ref{fig:dnr_resolution} in comparison to Fig.~\ref{fig:distance_error_histograms}). It shows an even greater dependence on shower energy, because the differences in viewing angles are larger between DnR signals than between channels. This makes the aforementioned effects even more significant, so that DnR pulses only become relevant at higher energies. One should keep in mind, however, that these efficiencies do not include the efficiency of identifying a neutrino event in the first place. Also, the phased array trigger is approximated with a simple threshold and some aspects of RNO-G, e.g. the noise level, are not yet confirmed by data at the time of writing.

\section{Electric field reconstruction}
\label{sec:efield}
Next to the vertex distance, we need to reconstruct the electric field of the radio signal from the recorded voltage waveforms. The Vpol and Hpol antennas are almost exclusively sensitive to the $\vec{e}_\theta$ or $\vec{e}_\phi$ components of the electric field, so in principle one could simply divide the spectrum of the voltages by the response of antenna and amplifier in the frequency domain to obtain the spectrum of the electric field. Unfortunately, this method requires signals to have a large SNR. Especially if the antenna has a relatively low sensitivity, as is the case for the Hpol antennas, thermal noise can be amplified and distort the result.

Instead, we use a method based on \emph{Information Field Theory} \cite{ift1,ift2}, which allows us to reconstruct even low-SNR events by including prior knowledge about the properties of the radio signal. This method is described in detail in \cite{Welling:2021cgl}, where RNO-G is also given as an application example. Therefore we will only give a brief description of the method here.

\subsection{Electric field reconstruction method }

The problem of electric field reconstruction can be thought of as determining $P(\mathcal{E}|U)$, the probability of the radio signal having a given frequency spectrum $\mathcal{E}$, given the measured voltages $U$, defined in the time domain, and finding the most likely electric field. Using Bayes' theorem, this probability can be calculated using
\begin{equation}
    P(\mathcal{E}|U) \propto P(\mathcal{E}) P(U|\mathcal{E}),
    \label{eq:IFT_bayes}
\end{equation}
if the prior probability $P(\mathcal{E})$ for the spectrum and the probability $P(U|\mathcal{E})$ for the measured voltages given a specific frequency spectrum are known.

We split the frequency spectrum into its absolute value and a phase 
\begin{equation}
    \label{eq:efield_model}
    \mathcal{E}(f) = E(f) \cdot \exp(i \cdot \varphi(f))
\end{equation}
and model each component separately.

The absolute value of the frequency spectrum is modeled using a generative process 
\begin{equation}
\label{eq:spectrum_model}
s(f)=\lg(E(f))
\end{equation}

$s(f)$ follows Gaussian statistics, with a correlation structure which is defined by its eigen-spectrum $\tau(k)$, which follows a power-law $\tau \propto k^{-\alpha}$, where $k$ is the wave number\footnote{Technically, since $E(f)$ is defined in the frequency domain, $k$ is defined in the time domain. We will ignore this subtlety here and use the notation that is most commonly used to describe the correlation structure.}, though small deviations from that are allowed. The slope $\alpha /2$ determines the smoothness of the electric field spectrum: A large value for $\alpha$ means only small $k$ in the eigenspectrum of $s(f)$, leading so a very smooth $s(f)$, which becomes more and more jagged with decreasing $\alpha$, as larger $k$ become more likely. We choose a value for $\alpha / 2$ that results in a rather smooth electric field spectrum, but leave the parameter some freedom to vary, should the data suggest a more jagged spectrum. With this, we can assign a prior probability to any $E(f)$, because the probability distribution of $s(f)$ is defined.

The phase is modeled by a linear function
\begin{equation}
    \label{eq:phase_model}
    \varphi(f) = \varphi_0 + mf
\end{equation}
where $\varphi_0$ and $m$ are Gaussian random variables. Such a phase leads to a single, short pulse, as is expected from an Askaryan signal. The slope $m$ corresponds to a time shift in the time domain and is set based on the previously determined signal timing. It is, however, given some room to vary within a few nanoseconds, to correct for possible errors in determining the pulse time.

Finally, the polarization of the radio signal is described by a parameter $\phi_{\text{pol}}$ called \emph{polarization angle}, so that
\begin{equation}
    \label{eq:polarization_model}
    \vec{\mathcal{E}}(f) = \mathcal{E}(f)\cdot \cos(\phi_{\text{pol}})\cdot\vec{e}_\theta + \mathcal{E}(f)\cdot \sin(\phi_{\text{pol}})\cdot\vec{e}_\phi
\end{equation}
The polarization angle is also modeled as a Gaussian random variable, but with a standard distribution chosen large enough to only have a negligible effect on the posterior distribution.

From equations \ref{eq:efield_model}-\ref{eq:polarization_model}, we can construct an operator that maps $s$, $\varphi_0$, $m$ and $\phi_{pol}$ onto $\mathcal{E}(f)$. This operator is invertible, which allows us to assign a prior probability to any $\mathcal{E}(f)$.
To do so, we calculate the values for $s(f)$, $\varphi_0$, $m$ and $\phi_{pol}$ that lead to the given $\mathcal{E}$. Because the prior probabilities for these variables are known, and assumed to be independent of each other, we can calculate their joint probability 
\begin{equation}
\begin{split}
P(\mathcal{E})&=P(s, \varphi_0, m, \phi_{pol}) \\
&=P(s)\cdot P(\varphi_0)\cdot P(m)\cdot P(\phi_{pol})
\end{split}
\end{equation}
as the product of the individual probabilities.

The antenna and amplifier responses are applied to the electric field, which can be expressed in the frequency domain by a diagonal matrix $\mathcal{D}$. Then the result is translated into the time domain via a Fourier transformation $\mathcal{F}$ to obtain the expected voltage waveform $U(t)=\mathcal{FDE}(f)$.
Neglecting uncertainties on the detector description and signal model, the difference between the expected and measured waveforms must be due to the noise N:
\begin{equation}
\begin{aligned}
    P(U|\mathcal{E})=P(N = U - \mathcal{FDE}) \\
    =\mathcal{N}(U-\mathcal{FDE}|0, \sigma_N)
\end{aligned}
\end{equation}
where the last term expresses the assumption that the noise is drawn from a Gaussian distribution centered around \SI{0}{V} with a standard deviation of $\sigma_N$, which can be determined from forced trigger data. With this, the maximum of Eq.~\ref{eq:IFT_bayes} can be determined using Metric Gaussian Variational Inference \cite{knollmueller2020metric}.

A strength of this method is that it only makes some general and well-founded assumptions about the properties of the radio signal. Considering that no experiment has yet detected a neutrino via its Askaryan emission, not relying on any physical model for the electric-field reconstruction is an advantage in itself, as it allows us to verify emission models by comparing the reconstructed electric field spectra to predictions. 

The reconstruction method requires the radio signal received by all channels to be very similar, which is the case if the antennas are close enough together. How much the radio signal differs between channels depends very much on the event geometry, which is not known at this point, except for the vertex position. Thus, there are three groups of channels that can be used to reconstruct the electric field, assuming they detected a radio signal: The phased array channels along with the two Hpol antennas directly on top of them, and the three channels at the bottom of each helper string. In principle, the electric field could also be reconstructed for the other channels on the \emph{power string}, but since they do not have any nearby Hpol antennas, it would have to be limited to the $\vec{e}_{\theta}$ component. It is also unlikely that a single channel will yield a better reconstruction than a group of three or six, so these channels are not used for the electric field reconstruction.

\subsection{Performance of the electric field reconstruction}

To asses the performance of the electric field reconstruction, we continue from the vertex reconstruction in Sec.~\ref{sec:vertex_reco_quality}.

In order to reconstruct the electric field of the radio signal, it is first necessary to identify which channels actually recorded a signal with a sufficiently high SNR. Only a tiny fraction of the radio signals that trigger RNO-G are actually from neutrinos, while the vast majority will come from thermal fluctuations or other radio sources in the environment. How to identify the few genuine neutrino events is its own topic, so here we assume that some method to do so is available. We can safely assume this is the case, as a proper event identification is necessary before an energy reconstruction could even start. 
We already used SNR $=2.5$ as the analysis threshold, so we assume that if a pulse has SNR $\geq 2.5$ in any channel, it can be identified and we just use the true pulse position as input. If a pulse is found, a region of \SI{\pm50}{ns} around it is marked as a \emph{signal region}. It should be noted that this only serves to mark the relevant section of the waveforms for the electric field reconstruction and no further information from this pulse finder is used. Any pulse with a lower SNR is classified as not found. As before, the maxima and minima are calculated from the noiseless waveforms for the purpose of calculating the SNR.

Next, we need to identify the directions from which the radio signals are received at each antenna. Using the reconstructed vertex position, we can calculate the expected signal propagation times for each ray path. With the template correlation method used for the vertex reconstruction, the signal arrival time is determined and compared to the expectation for each ray tracing solution to identify the most likely ray path. Once the correct ray tracing solution is found, the signal arrival direction and time shift relative to the other antennas can also be determined. This is done for every signal region. However, for most events, the Hpol channels will not detect a signal with a sufficiently high SNR to identify a signal region. In order to be able to use them for the electric field reconstruction, a linear function is fitted to the arrival times and zenith angles\footnote{The azimuth is the same for all antennas on the same string.} as a function of depth of the Vpol antennas on the same string as the Hpol. This function is then used to extrapolate these values for the Hpol antennas. While this approach assumes a planar wave front, it is accurate enough for antennas that are spaced as closely as in this case.

\begin{figure}
    \centering
    \includegraphics[width=.45\textwidth]{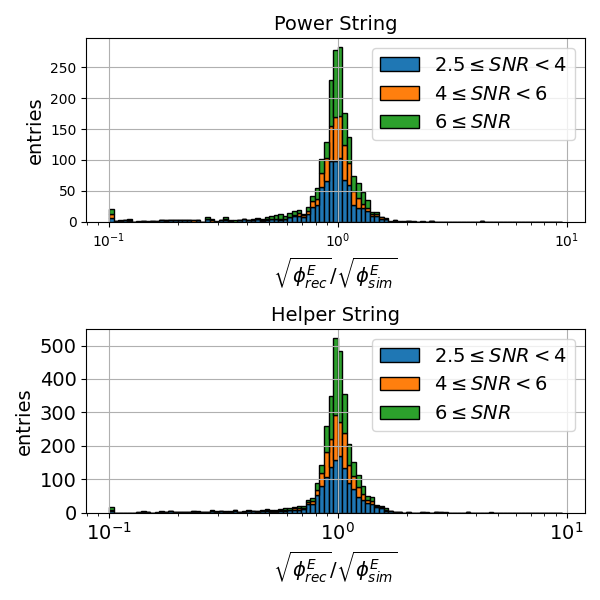}
    \caption{Ratio between the square root of the energy fluence of the reconstructed and the actual radio signal detected at the power string (top) and the two helper strings (bottom). Results are divided into categories by the maximum signal-to-noise ratio of the channels used for the reconstruction. The outermost bins are overflow bins.}
    \label{fig:fluence_hist}
\end{figure}

As discussed above there are three groups of channels for which the electric field can be reconstructed: The phased array along with the two Hpol antennas directly above it, and the antennas on each of the helper strings. In principle, the phased array has twice the number of channels and is, thus, most promising to use for reconstructing the radio signal. However, depending on the event geometry, the SNR can vary greatly between strings. Therefore, we reconstruct the electric field for all channel groups (provided a \emph{signal region} was found).

The performance of the electric field reconstruction with regards to the energy fluence $\phi^E$ of the radio signal is shown in Fig.~\ref{fig:fluence_hist}. Except for a few outliers, which were shown to be easily identifiable in \cite{Welling:2021cgl}, the energy fluence is well-reconstructed, with the 68\% quantiles at about 20\% or better, even for low-SNR events, as shown in Tab.~\ref{tab:electric_field_resolution}. Interestingly, the reconstructions using \emph{helper strings} perform roughly as well as using the \emph{power string}, even though only half as many channels are available.

\begin{figure}
    \centering
    \includegraphics[width=.45\textwidth]{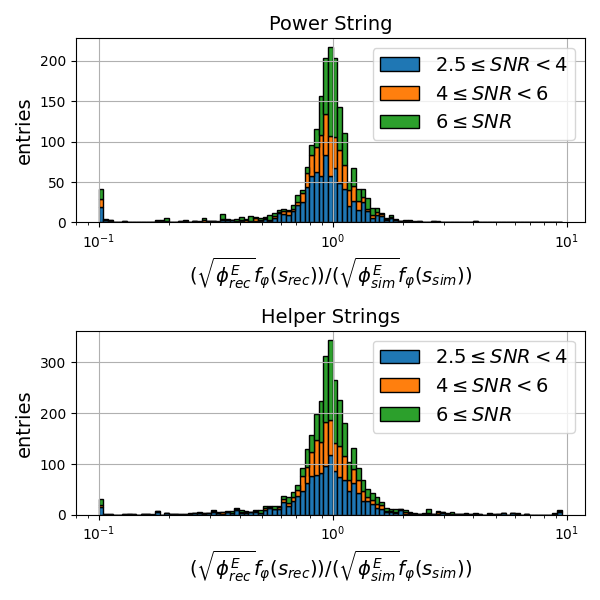}
    \caption{Ratio of the square root of the energy fluence multiplied with the correction factor $f_\varphi(s)$ for the viewing angle defined in Eq.~\ref{eq:spectrum_shape_formula} calculated from the reconstructed and the actual radio signal at the power string (top) and each of the two helper strings (bottom). For events where $s>10$, $f_\varphi(s^\prime)$ was used instead. Events are divided into categories by the maximum signal-to-noise ratio of the channels used in the reconstruction. The outermost bins are overflow bins.}
    \label{fig:energy_factor}
\end{figure}

In addition to the energy fluence, the shape of the electric field spectrum, expressed by the parameter $s$ (see Eq.~\ref{shape_parameter_definition}), is also needed to correct for the effect of the viewing angle (see Eq.~\ref{eq:spectrum_shape_formula}). Fig.~\ref{fig:energy_factor} shows the ratio of the energy estimator $\sqrt{\phi^E} f_\varphi (s)$ calculated from the reconstructed and from the true electric field spectrum. To reconstruct the shower energy, these still have to be corrected for propagation effects, which will be done in the next section. Because the attenuation length in ice is frequency dependent, correcting for attenuation will change the shape of the frequency spectrum as well, but this effect is small. Therefore, Fig.~\ref{fig:energy_factor} gives an estimate of the uncertainty on the shower energy due to the electric field reconstruction. Looking at the 68\% quantiles of the energy estimator compared to the energy fluence alone (see Tab.~\ref{tab:electric_field_resolution}), the viewing angle correction $f_\varphi(s)$ increases the total uncertainty relatively little. This is because the expression in Eq.~\ref{eq:spectrum_shape_formula} is a function of the logarithm of $s$, so changes to $s$ have a smaller impact.

\begin{table}[]
    \centering
    \begin{tabular}{|c|c|c|c|}
        \hline
         & $2.5\leq$ SNR $< 4$& $4\leq$ SNR $< 6$ & $6\leq$ SNR\\
         \hline
         $\sqrt{\phi^E}$& \multicolumn{3}{c|}{Power String} \\
         \hline
         median & 0.98 & 0.96 & 0.99 \\
         $\sigma_{68}$ & [0.79, 1.17] & [0.79, 1.08] & [0.85, 1.14] \\
         \hline
          $\sqrt{\phi^E}f_\varphi(s)$ & \multicolumn{3}{c|}{} \\
         \hline
         median & 0.94 & 0.93 & 0.98 \\
         $\sigma_{68}$ & [0.71, 1.17] & [0.74, 1.08] & [0.76, 1.18] \\
         \hline
         $\sqrt{\phi^E}$ & \multicolumn{3}{c|}{Helper Strings} \\
         \hline
         median & 1.02 & 1.00 & 1.01 \\
         $\sigma_{68}$ & [0.85, 1.17] & [0.87, 1.10] & [0.88, 0.14] \\
         \hline
          $\sqrt{\phi^E}f_\varphi(s)$ & \multicolumn{3}{c|}{} \\
         \hline
         median & 0.98 & 0.97 & 1.01 \\
         $\sigma_{68}$ & [0.73, 1.33] & [0.74, 1.28] & [0.85, 1.26] \\
         \hline
    \end{tabular}
    \caption{Median and 68\% quantiles $\sigma_{68}$ of the ratios between reconstructed and actual square root of the energy fluence shown in Fig.~\ref{fig:fluence_hist}, as well as between the energy estimators $\sqrt{\phi^E}f_\varphi(s)$ calculated from the reconstructed and the actual electric field shown in Fig.~\ref{fig:energy_factor}. Shown are results for the channel groups at the bottom of the power string and of the helper strings, as well as for different maximum SNRs.}
    \label{tab:electric_field_resolution}
\end{table}

\section{Shower energy reconstruction}
\label{sec:shower_energy}
All information can now be combined to reconstruct the shower energy. First, a correction for the attenuation of the radio signal on its way from the shower to the station is needed. With the location of the shower known, the ray tracing can be redone to determine the path of the radio signal. Because the attenuation length changes with depth and frequency, the path loss $\mathcal{L}(f)$ is defined in the frequency domain using
\begin{equation}
    \mathcal{L}(f) = \exp \left( -\int \frac{1}{l_{\text{att}}(\vec{x},f)} ds \right)
\end{equation}
by integrating over the ray path. $l_{\text{att}}(\vec{x}, f)$ is the depth- and frequency-dependent attenuation length. With this, we calculate the frequency spectrum of the electric field corrected for propagation effects as 
\begin{equation}
    \mathcal{E}_0(f) = \mathcal{E}(f) / \mathcal{L}(f) / \left( \frac{l}{l_{\text{ref}}} \right)
\end{equation}
where $l$ is the length of the ray path from the interaction vertex to the detector and $l_{\text{ref}}$ the reference distance for which the parametrization Eq.~\ref{eq:spectrum_shape_formula} was calculated, in our case \SI{1}{km}.

After these corrections the shower energy can be calculated from Eq.~\ref{eq:spectrum_shape_formula}. Reconstructed electric fields are available from all three strings for 60\% of events, while for 22\% of the cases the reconstruction can use two of the strings. Just for 18\% of the events the electric field can be reconstructed for one string only. If more than one field reconstruction is available, the shower energy is calculated per electric field and the final result is calculated as the mean of the energies.

\subsection{Performance of the shower energy reconstruction}
\begin{figure}
    \centering
    \includegraphics[width=.45\textwidth]{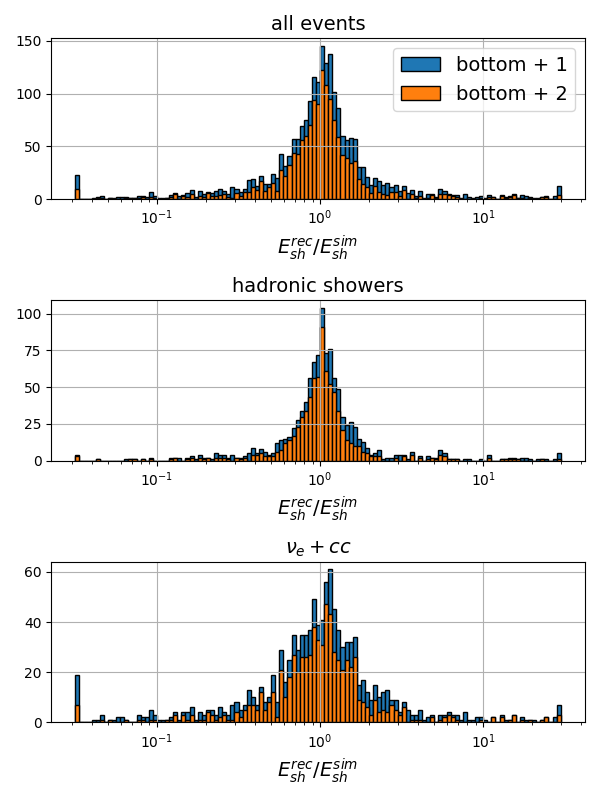}
    \caption{Ratio between reconstructed and actual shower energy for all events (top), hadronic showers (middle) and charged-current interactions of electron neutrinos (bottom). Shown are results of events where at least one of the channels at \SI{\sim 100}{m} as well as at least one (blue) or at least two (orange) of the other channels on the \emph{power string} detected a signal with SNR$\geq 2.5$. The outermost bins are overflow bins.}
    \label{fig:energy_reco_hist}
\end{figure}

\begin{figure}
    \centering
    \includegraphics[width=.45\textwidth]{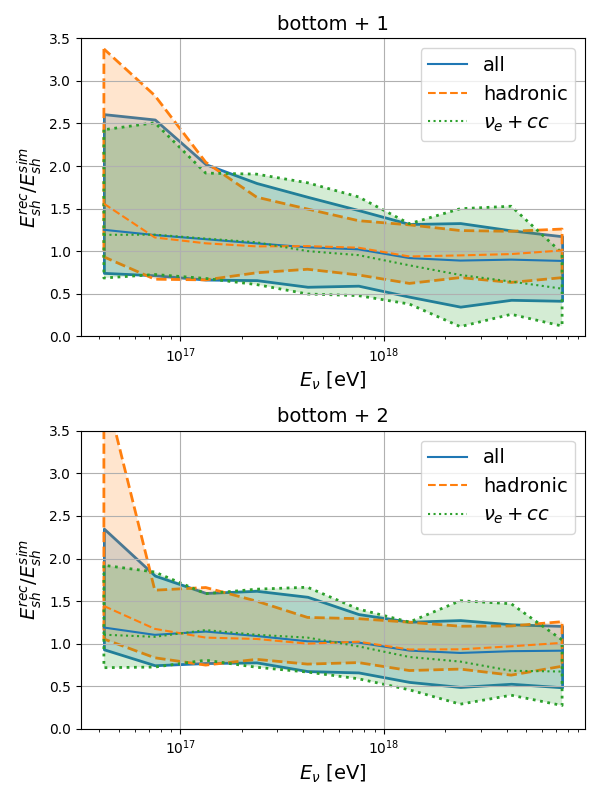}
    \caption{Median and 68\% quantiles of the ratios between reconstructed and actual shower energies as function of neutrino energy. The results are shown for events that only produced a hadronic showers, as well as for charged-current interactions of electron neutrinos. The top panel shows events where at least one of the channels at \SI{\sim 100}{m} detected a pulse with SNR $\geq 2.5$ as well as at least one of the other channels on the \emph{power string}, the bottom panel shows results with at least two additional channels on the \emph{power string}.}
    \label{fig:energy_reco_quantiles}
\end{figure}

\begin{figure}
    \centering
    \includegraphics[width=.45\textwidth]{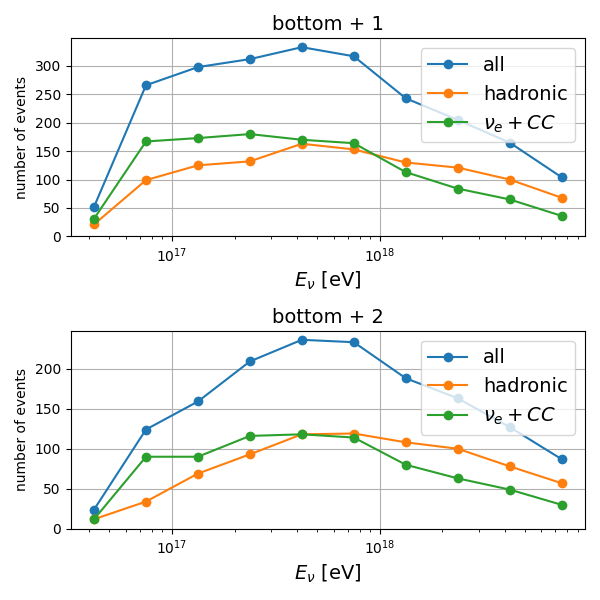}
    \caption{Number of events contained in each of the energy bins in Fig. \ref{fig:energy_reco_quantiles}. Events are divided into the same categories as explained in Fig. \ref{fig:energy_reco_quantiles}}
    \label{fig:number_of_events}
\end{figure}

The results of the reconstruction of the shower energy are shown in Fig.~\ref{fig:energy_reco_hist}. An energy resolution of better than a factor of 2 is achieved, if at least one of the channels of the \emph{power string} above the phased array detected a pulse of $\text{SNR}\geq 2.5$ and improves further if at least 2 additional channels are available. Unsurprisingly, the energy reconstruction works better for hadronic showers than for events from charged-current interactions of electron neutrinos, where potentially interfering hadronic and electromagnetic shower are present. The 68\% quantiles on the $\lg(E_{\text{sh}}^{\text{rec}}/E_{\text{sh}}^{\text{sim}})$ is $[-0.15, 0.18]$ with the 1 channel cut and $[-0.13, 0.12]$ with the 2 channel cut. On a linear scale the 68\% quantiles of $E_{\text{sh}}^{\text{rec}}/E_{\text{sh}}^{\text{sim}}$ are $[0.58, 1.66]$ and $[0.65, 1.49]$, respectively. The uncertainty is smaller for events that only produce a hadronic shower, with a 68\% quantiles of $[0.70, 1.54]$ and $[0.74, 1.34]$, while for events that produce both a hadronic and an electromagnetic shower it is $[0.47, 1.86]$ and $[0.57, 1.63]$ for the 1 channel and the 2 channels cuts, respectively.

The energy resolution as a function of neutrino energy is shown in Fig.~\ref{fig:energy_reco_quantiles}, along with Fig. \ref{fig:number_of_events} showing the statistical power of each energy bin. For hadronic showers, the relative uncertainty decreases with increasing energy, which is not the case for the events with electromagnetic showers. Especially if at least 2 of the channels above the phased array have a signal with SNR $\geq 2.5$, the reconstruction tends to underestimate for shower energies above \SI{\sim 1e18}{eV}. This is roughly the energy where the LPM effect starts to have a strong impact on the radio signal from the electromagnetic shower. This effect is already visible in Fig.~\ref{fig:viewng_angle_parametrization}, but another effect is relevant here as well: Because the electromagnetic shower is elongated, but the hadronic one is not, their maxima may be seen at different viewing angles, so that only one of the radio signals is detected, while the other falls below the noise level. In that case, we are practically only reconstructing the energy of one of the subshowers. If the energy further increases, the electromagnetic shower will consist of multiple spatially displaced sub showers \cite{Glaser:2019cws}. Then, only the energy of one of the sub showers might be reconstructed which further complicates the reconstruction. As we are still defining the true shower energy as being equal to the neutrino energy for these events, this causes the $E_{\text{sh}}^{\text{rec}}/E_{\text{sh}}^{\text{sim}}$ to decrease for these events.

\subsection{Influence of systematic uncertainties}
So far, we have assumed a perfectly calibrated detector without any experimental uncertainties except for interference with thermal noise. This is not achievable in reality. At the time of writing, RNO-G is in its first deployment season, which of course means that no in-situ calibration data is available. A thorough estimation of systematic uncertainties will only be possible later when the performance of RNO-G is tested in the field. Instead, we will explore common sources of systematic uncertainties to estimate what uncertainties on the detector performance would be acceptable for the energy resolution to remain useful. In particular, the three most important uncertainties are the antenna response, timing uncertainties between different channels, and the index of refraction of the ice.

The antenna responses for RNO-G have been carefully simulated, but in-situ calibration of the antennas after deployment is challenging. Calibration campaigns to verify the simulated antenna responses are planned for future deployment seasons, but some uncertainty is unavoidable. The same is true for the response of the amplifiers or any other element of the signal chain. However, as these have been calibrated in the lab before deployment, mimicking the situation in the field, the antenna uncertainties are expected to be larger than the uncertainties on the amplifier response. By assuming uncertainties on the antenna response and investigating the effect it has on the energy reconstruction, we can estimate the required accuracy with which the antennas will have to be calibrated. We will consider two primary ways for the antenna response to affect the result: A wrong overall amplitude, or a wrong shape of the antenna response.

\begin{figure}
    \centering
    \includegraphics[width=.45\textwidth]{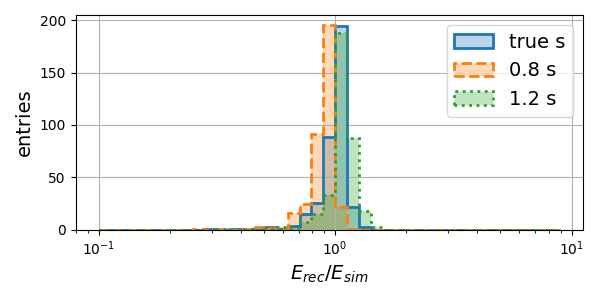}
    \caption{The estimated shower energy according to Eq.~\ref{eq:spectrum_shape_formula}, calculated for the radio signals from simulated hadronic showers using the true $s$ parameter (blue, solid) and an $s$ parameter that is under- (orange) or overestimated (green) by 20\%.}
    \label{fig:slope_error}
\end{figure}

If the overall amplitude of the antenna response is wrong, the consequences are straightforward: $\sqrt{\phi^E_{\text{rec}}}$ is proportional to the amplitude of the antenna response, and also proportional to the energy estimator. This means the relative uncertainty on the amplitude of the antenna response simply propagates linearly onto the reconstructed energy. More problematic are differences in overall scale between antennas. However, since great care has been taken to produce uniform antennas, we will not discuss this uncertainty at this point without in-field data to suggest large differences. Other radio-based detectors have demonstrated antenna calibrations at the 10\% level \cite{PierreAuger:2017xgp}, so we do expect to be able to obtain a similar performance. It should also be noted that systematic differences between Hpol and Vpol antennas can introduce uncertainties in the reconstruction of the electric field, similarly to an overall amplitude mismatch. 

An incorrect shape of the antenna response is hard to quantify, but can be assessed qualitatively. In \cite{Welling:2021cgl} it is shown that Information Field Theory can still reconstruct the electric field if there are uncertainties on the shape of the antenna response, as well as uncertainties on the phase of the antenna and signal chain, but the same difference will show up in the shape of the reconstructed electric field spectrum. This means that as a consequence the reconstructed slope parameter $s$ (see Eq.~\ref{shape_parameter_definition}) would be affected. To see what an uncertainty on $s$ would mean for the reconstructed energy, we generate radio signals from simulated hadronic showers, in the same way as was done to create Fig.~\ref{fig:viewng_angle_parametrization}. Then we determine $\Phi_E$ and $s$ and use it to estimate the energy using Eq.~\ref{eq:spectrum_shape_formula}, but after applying an error to $s$. Even assuming a 20\% uncertainty on $s$ from the antenna response, Fig.~\ref{fig:slope_error} shows that it has only a very small influence on the reconstructed shower energies, only shifting the mean of the $E_{\text{rec}}/E_{\text{sim}}$ distributions by -0.05 and 0.04 compared to using the true $s$. While the uncertainties on the antenna responses that RNO-G will have is unknown at this point, these results provide guidance to what precision the detector needs to be calibrated.

\begin{figure}
    \centering
    \includegraphics[width=.45\textwidth]{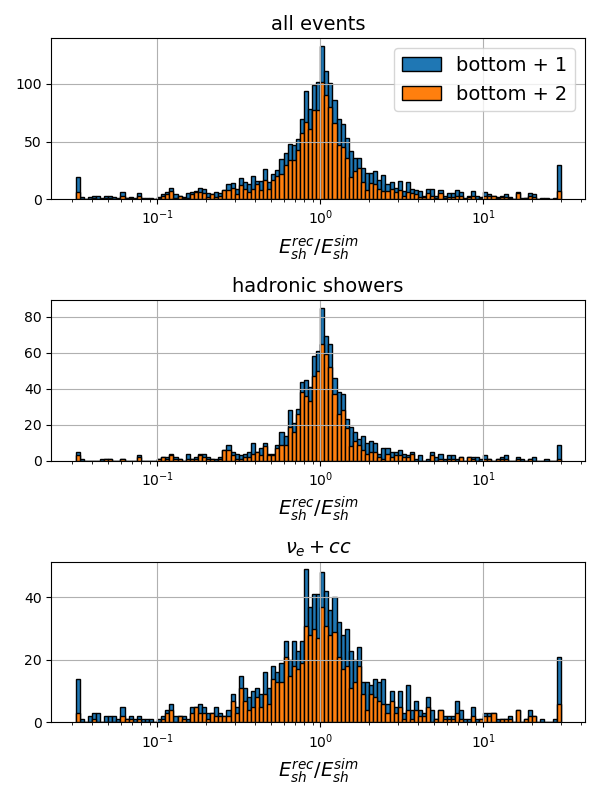}
    \caption{Ratio between reconstructed and actual shower energy if a random time offset drawn from a normal distribution with $\sigma_t=0.3ns$ is applied to each channel. Shown are results of events where at least one of the channels at \SI{\sim 100}{m} as well as at least one (blue) or at least two (orange) of the other channels on the \emph{power string} detected a signal with SNR$\geq 2.5$. The outermost bins are overflow bins.}
    \label{fig:energy_reco_hist_time_jitter}
\end{figure}
Uncertainties on the timing can be caused by delays in the signal chain that are unaccounted for. These can be reduced by careful calibration before deployment and are expected to be very stable in time. A more severe cause of uncertainties is due to the position of the antennas, either because of the leeway on their exact position within their hole or because of some tilt of the hole itself. This would have the same effect as an uncertainty on the timing, because the radio signal arrives at a different time from what is expected for the assumed antenna positions. To simulate this we shift the waveforms of each channel by a random offset drawn from a normal distribution with a standard deviation of $\sigma_t=0.3$ ns and redo the energy reconstruction. The ARA collaboration has shown a timing calibration to around \SI{0.1}{ns} or better \cite{ARA_ICRC}, so this is a rather pessimistic assumption. Comparing the resulting energy reconstruction (Fig.~\ref{fig:energy_reco_hist_time_jitter}) to Fig.~\ref{fig:energy_reco_hist} we see that the number of outliers is increased, especially if only one of the channels above the phased array detected a radio signal. Most events still remain little affected and the 68\% quantiles on $\lg(E_{\text{sh}}^{\text{rec}}/E_{\text{sh}}^{sim})$ only grow from $[-0.16, 0.18]$ to $[-0.19, 0.20]$ or from $[-0.13, 0.12]$ to $[-0.17, 0.14]$ with the 1 or 2 channel cut, respectively.

\begin{figure}
    \centering
    \includegraphics[width=.45\textwidth]{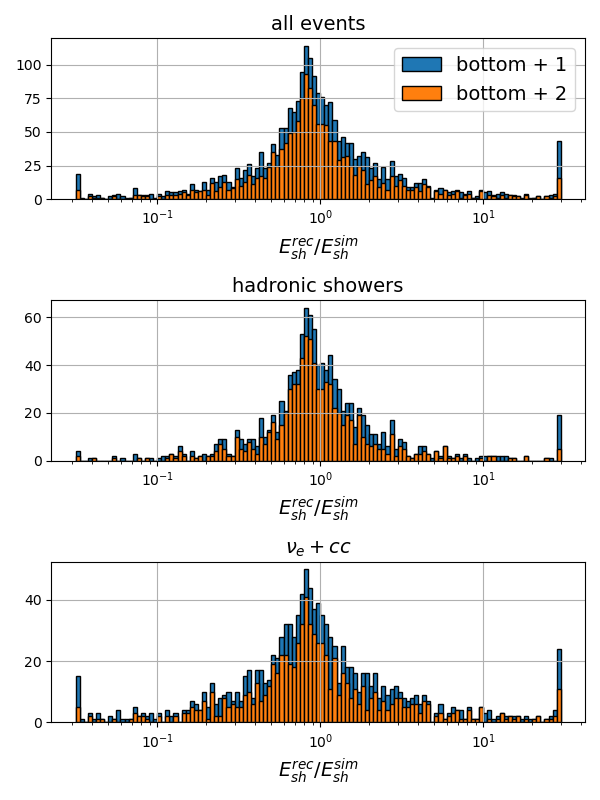}
    \caption{Ratio between reconstructed and actual shower energy if the index of refraction models used for simulation and reconstruction differ. Shown are results of events where at least one of the channels at \SI{\sim 100}{m} as well as at least one (blue) or at least two (orange) of the other channels on the \emph{power string} detected a signal with SNR $\geq 2.5$. The outermost bins are overflow bins.}
    \label{fig:energy_reco_hist_ice_model}
\end{figure}

Finally, we consider uncertainties on the index of refraction of the ice. A profile of the index of refraction as a function of depth of the ice near Summit Station has been developed \cite{Deaconu_2018bkf}, but only to a depth of \SI{120}{m}. In the future, dedicated measuring campaigns will develop a profile for the entire detection volume, as it has been done in other places. A compilation of index of refraction measurements at various places in Antarctica can be found at \cite{Barwick_2018} along with the parameters for the parametrization in Eq.~\ref{eq:index_of_refraction}. To assess the effect that uncertainties of these measurements have on the energy reconstruction, we rerun the Monte Carlo simulations, but change the $\Delta_z$ parameter of the index of refraction profile from the original value of $\Delta_z=0.51$ to $\Delta_z=0.50$ and the $z_0$ parameter from \SI{37.25}{m} to \SI{36.25}{m}. We then rerun the entire energy reconstruction assuming the original values of $\Delta_z$ and $z_0$. Both of these assumed uncertainties are similar to those shown in \cite{Barwick_2018}.

Fig.~\ref{fig:energy_reco_hist_ice_model} shows the impact that ice model uncertainties have on the reconstructed energy, resulting in 68\% quantiles of $[-0.29, 0.28]$ and $[-0.27, 0.23]$ on $\lg(E_r/E_s)$ for the 1 channel and 2 channel cuts, respectively. This means uncertainties on the index of refraction profile of the ice can have a significant impact on the energy reconstruction and careful calibration measurements will be required. For completeness, it should be mentioned that this study does not yet account for other potential ice effects such as birefringence or layers in the ice, as discussed in \cite{ICRC_RadioPropa}. However, information about these effects on radio emission in the ice at Greenland are currently not quantifiable beyond speculation, which is why we chose to exclude them at this point. 

\section{Neutrino energy}
\label{sec:inelasticity}
\begin{figure}
    \centering
    \includegraphics[width=.45\textwidth]{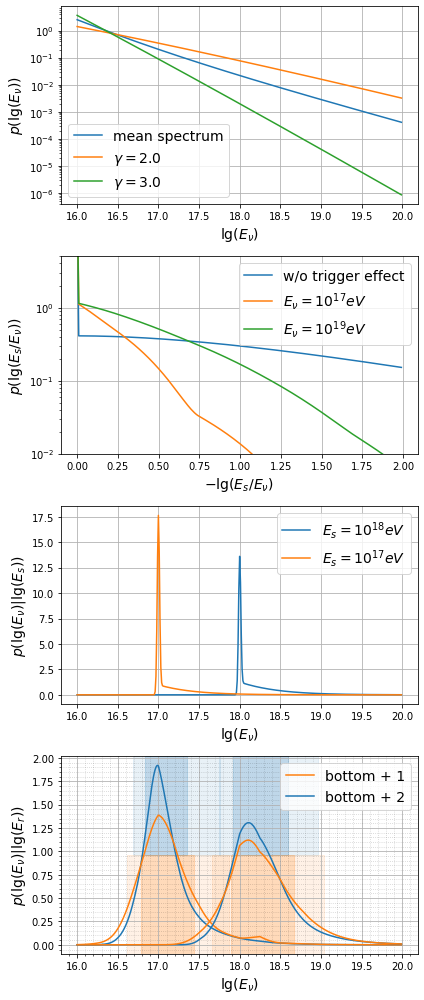}
    \caption{Top: Energy distribution of neutrinos interacting in the detector volume assuming a power law spectrum with specific indices (green, orange) and over a distribution of spectra (blue). Second row: Distribution of the fraction of the neutrino energy that goes into the shower for neutrino-nucleon interactions (blue) and the distribution when trigger efficiencies are included (orange, green). Third row: Probability for a shower with a given energy to have been generated by a neutrino with energy $E_\nu$. Bottom: The same distribution as above, but assuming an uncertainty on the reconstructed shower energy. The shaded regions mark the 68\% and 90\% quantiles of the distributions.}
\label{fig:inelasticity}
\end{figure}
If a neutrino interacts in the ice, only a certain fraction of its energy will be transferred to the nucleon it interacted with, while the rest remains with the neutrino (in the case of neutral current interactions) or the charged lepton created (in the case of charged current interactions). This fraction is randomly distributed and can typically not be measured for a given event, so it represents an irreducible uncertainty on the achievable energy resolution of any neutrino detector. In the case of an electron neutrino interacting via charged current interaction (referred to as a $\nu_e$+CC event), the electron initiates a shower as well, so that the entire neutrino energy ends up in the two showers. Measuring both showers results in a reduction of the uncertainty on the neutrino energy, which could exploited if such events could be uniquely identified. So far, no method to do so has been developed, so we will not discuss this option here. Instead, we treat all cases together where a shower energy has been reconstructed as described above. 

We can use Bayes' theorem to estimate the probability $p(E_\nu|E_r)$ for the neutrino to have an energy $E_\nu$ given that the energy of the shower was reconstructed to be $E_r$. Since the energies in question are distributed over several orders of magnitude, it is convenient to think about the problem in terms of the base 10 logarithm $\lg(E)$ and we thus get:
\begin{equation}
    \label{eq:inelasticity_bayes}
    p(\lg(E_\nu)|\lg(E_{r})) = \frac{p(\lg(E_{r})|\lg(E_\nu))\cdot p(\lg(E_\nu))}{p(\lg(E_{r}))}
\end{equation}
If we know that neutrinos follow a given spectrum $S$, the energy distribution of neutrinos interacting in the detector volume is given by
\begin{equation}
    p(\lg(E_\nu)|S) = \frac{\Phi_\nu (E_\nu) / \lambda_\nu (E_\nu) \cdot E_\nu }{\int_{E_-}^{E_+} \Phi_\nu (E_\nu^\prime) / \lambda_\nu (E_\nu^\prime) \cdot E_\nu^\prime \cdot dE_\nu^\prime}
\end{equation}
where $\Phi_\nu$ is the neutrino flux and $\lambda_\nu$ is the interaction length. The $E_\nu$ term comes from the Jacobian when transforming from $E_\nu$ to $\lg(E_\nu)$. The problem now is that no neutrino has been detected at these energies yet, so we cannot know the spectrum. Instead, we have to consider a range of spectra $S$, to each of which we assign a prior probability $p(S)$. Then we can calculate the expected neutrino energy spectrum using:
\begin{equation}
    p(\lg(E_\nu)) = \int_{S_-}^{S_+} p(\lg(E_\nu)|S)\cdot p(S) dS
\end{equation}
What spectra are considered possible or even probable has to be based on our prior knowledge and is of course very subjective. For our calculations, here, we assume a power law spectrum
\begin{equation}
    \Phi_\nu (E_\nu) \propto E_\nu^{- \gamma}
\end{equation}
where the spectral index is equally likely for all values $2 < \gamma < 3$. The resulting neutrino energy distribution is shown in Fig.~\ref{fig:inelasticity}, (top). 

It is of course possible that a single power-law cannot describe the neutrino flux, especially across the entire energy range. It seems probable, for example, that a power-law from an astrophysical flux at PeV energies will be dominated by a cosmogenic flux \cite{vanVliet:2019nse}, not following a power-law, at EeV energies. 

For the sake of clarity, we will only discuss generic power law spectra here, as extrapolation of the neutrino flux detected by IceCube. Still, we have tested a cosmogenic neutrino flux and even different spectral indices (see \cite{Christoph_ICRC_2021} for details) and have found the results to be very similar. The code used for these calculations is available online \cite{inelasticity_jupyter} and allows one to test other spectral models. 
We also note that ultimately, the goal is to obtain a neutrino energy spectrum rather than a single event energy. For this an adapted, iterative approach will certainly be useful. 

For events that only produce a hadronic shower, the fraction of the neutrino energy that goes into the shower is given by the inelasticity $y$. 
It should be mentioned that the inelasticity distribution carries uncertainties. The inelasticity has only been measured experimentally to TeV energies \cite{PhysRevD.99.032004} and at even higher energies, aspects such as nuclear modifications to neutrino cross-sections \cite{PhysRevC.102.015808} or electromagnetic contributions, subdominant to deep-inelastic scattering \cite{PhysRevLett.80.900,PhysRevD.101.036010} may become relevant. We ignore the uncertainties for the purpose of this paper and use the inelasticity distribution as implemented in NuRadioMC \cite{Glaser:2019cws} based on \cite{CooperSarkar:2011pa,Connolly:2011vc,Gandhi:1995tf}.

If an electron neutrino undergoes charged current interaction, all the energy ends up in the two showers produced, so that effectively $E_s=E_\nu$.
For a neutrino event whose flavor and interaction type we do not know, we can combine these two scenarios with
\begin{equation}
    p(\lg(\frac{E_s}{E_\nu})) = p(\lg(\frac{E_s}{E_\nu})|h)\cdot P(h) + p(\lg(\frac{E_s}{E_\nu})|e)\cdot P(e)
\end{equation}
where $p(\lg(\frac{E_s}{E_\nu})|h) = p(\lg(y))$ is the $\frac{E_s}{E_\nu}$ distribution assuming only a hadronic shower was produced, which is equal to the inelasticity distribution, and $p(\lg(\frac{E_s}{E_\nu})|e) = \delta(\lg(\frac{E_s}{E_\nu}))$ assuming an electromagnetic shower was produced as well. The probability $P(e)=1-P(h)=\frac{1}{3}\cdot 0.71$ for this is the probability for the neutrino flavor being $\nu_e$, assuming equal mixing, times the probability for it to undergo a charged current interaction. One can simply update this probability in case additional information about the interaction type is reconstructed.

While this is the distribution for showers produced in the ice, it is not necessarily the distribution for the showers that are detected, because events with large $\frac{E_s}{E_\nu}$ have a higher chance of triggering the detector. Therefore the probability $p_T(E_s)$ for a shower with a given energy to trigger the detector needs to be included\footnote{To be precise, it would need to be the probability for the shower to trigger the detector, to be identified as a neutrino and to meet any quality criteria required for an energy reconstruction. The efficiency to identify neutrino events is not yet known, so we only use the trigger efficiency here.}. This trigger probability is different for the two event types, especially at higher energies when the radio signal is affected by the elongation of the electromagnetic shower from the LPM effect. Considering these effects, we get 
\begin{equation}
\begin{aligned}
    p(\lg(E_s)|\lg(E_\nu)) = \delta(\lg(\frac{E_s}{E_\nu}))\cdot p_T(E_s|e)\cdot P(e) +\\
    p \left( \lg(y)=\lg(\frac{E_s}{E_\nu})\right) \cdot p_T(E_s|h) \cdot P(h)
\end{aligned}
\end{equation}
Including the trigger efficiency shifts the $\frac{E_s}{E_\nu}$ distribution towards larger $\frac{E_s}{E_\nu}$ (see Fig.~\ref{fig:inelasticity}, second row). This effect is much stronger at low energies, where the trigger efficiency falls off faster towards lower shower energies.

At this point, we could calculate $p(\lg(E_\nu)|\lg(E_s))$ using the mean spectrum from Fig.~\ref{fig:inelasticity}, top and the $\frac{E_s}{E_\nu}$ distributions, to get the best possible neutrino energy resolution if we were able to reconstruct the shower energy perfectly (Fig.~\ref{fig:inelasticity}, third row). Properly quantifying this best-case energy resolution is made difficult by the delta distribution from the $\nu_e$+CC events, so we apply a small uncertainty on the shower energy of 2\% here. The uncertainty is small enough to conserve the shape of the $p(\lg(E_\nu)|\lg(E_s))$ distribution, but it removes the jump when integrating over the probability density. Thus, we can calculate the 68\% quantiles for the two example distributions, which are $[16.99, 17.08]$ and $[17.99, 18.21]$. Although the distribution is sharply peaked at the shower energy and falls off quickly, it has long tails. The reason for the higher peak and the smaller uncertainty quantiles at \SI{1e17}{eV} compared to \SI{1e18}{eV} is the reduced trigger efficiency of $\nu_e$+CC events at high energies. From this plot we see that even with a perfect shower energy reconstruction, the neutrino energy resolution would be limited.

The obtainable shower energy resolution is energy-dependent and also different for $\nu_e$ undergoing charged current interactions as shown in Fig.~\ref{fig:energy_reco_quantiles}. Similarly to that plot, we determine the median and the 68\% quantiles for $\lg(E_r/E_s)$  as a function of the shower energy. We approximate $p(\lg(E_r)|\lg(E_s))$ as a normal distribution $\mathcal{N}(\lg(E_r/E_s)|\mu_E, \sigma_E)$ and use the median as the offset $\mu_E$ and half the width of the 68\% quantiles as the standard deviation $\sigma_E$. We only determined the shower energy resolution for the energy range \SIrange{5e16}{1e19}{eV}, so for energies outside of this range, we use the resolutions from the \SI{5e16}{eV} and \SI{1e19}{eV} energy bins. This allows us to calculate the posterior probability of the reconstructed shower energy for a given neutrino energy as
\begin{equation}
\begin{aligned}
    p(\lg(E_r)|\lg(E_\nu))=\int \mathcal{N}(\lg(E_r/E_s)|\mu_E, \sigma_E)\cdot \\ 
    p(\lg(E_s)|\lg(E_\nu)) \; d\lg(E_s)
\end{aligned}
\end{equation}
which again has to be calculated separately for the two event types and then combined using $P(h)$ and $P(e)$.

Finally, the denominator of Eq.~\ref{eq:inelasticity_bayes} just serves to make sure it is properly normalized and can be determined by integrating the numerator over $\lg(E_\nu)$. As integration limits we use $16<\lg(E_\nu)<20$. The resulting neutrino energy resolution for the shower energy resolutions shown in Sec.~\ref{sec:shower_energy} is shown in Fig.~\ref{fig:inelasticity}  (bottom) for neutrino energies of \SI{1e17}{eV} and \SI{1e18}{eV}. Including the uncertainty on the shower energy changes the 68\% quantiles to $[16.83, 17.35]$ and $[17.92, 18.59]$ with the stricter quality cut of requiring at least 2 additional channels on the \emph{power string} or to $[16.79, 17.44]$ and $[17.89, 18.67]$ when requiring only one additional channel. 

The larger uncertainty at \SI{1e18}{eV} has two reasons, both caused by the LPM effect: The trigger probability for events that only produce a hadronic shower, and are therefore affected by the uncertainty on $\frac{E_s}{E_\nu}$, is larger and the uncertainty on the shower energy is larger for $\nu_e$+CC events at higher energies. To improve this, the most promising strategy is to improve the shower energy resolution for events from electron neutrinos undergoing charged current interactions and to develop a method to identify these events, as discussed in more detail in \cite{Christoph_ICRC_2021}.

It should be noted that in this analysis we have not considered any background, which can come from several sources for a radio experiment. Air showers from cosmic rays can produce radio signals, as they propagate through the air and when a not yet fully developed air shower reaches the ice surface \cite{deVries:2015oda}. Additionally, high-energy muons produced in an air shower can penetrate deep into the ice and produce showers there \cite{Garcia-Fernandez:2020dhb}. At this point, it has to be assumed that these would be difficult to distinguish from neutrino-induced showers. If detectable, all of these backgrounds would show a different spectrum as function of energy. The backgrounds can be mitigated by detecting the air shower using the upward-pointing antennas at the surface.

Other backgrounds are radio emissions from the environment around the detector. Windy weather conditions have been found to produce radio pulses, potentially because of triboelectric effects of the snow, which may mimic neutrino signals \cite{Barwick:2016mxm,Mikhailova:2021ccy}. However, a single, large pulse would be required for it to be mistaken for a neutrino signal, and the direction of the signal would likely reconstructed to the surface. This background can also be reduced by vetoing high-winds periods. Finally, human activity around the detection site inevitably produces radio emission. The nature and characteristics of this background are very difficult to predict and unique to each site. This will have to be investigated and subsequently modeled once RNO-G stations have gathered a sufficient amount of data in the field.

Including these background in our model at this point is very difficult. The background from air showers is not well-constrained yet, as for example models for the PeV muon flux have not been experimentally confirmed and rates of air shower emission depend on partly not well-known details of the ice. The veto efficiency of the surface antennas of RNO-G depends on the in-situ trigger performance, which is in-turn a function of the anthropogenic background. 
The background from the environment will have to be measured in situ and modeled throughout the year. Together with the large uncertainties on the expected neutrino flux, we can not yet give a good estimate of the purity of the data set that we expect to achieve. We therefore decided to ignore the background for now in the modeling of the energy resolution. This will have to be revisited once more information is available.

\section{Summary and conclusion}
We have presented a method to reconstruct the energy of neutrinos above above \SI{50}{PeV} using their radio signature in ice. The method has been developed for the Radio Neutrino Observatory in Greenland (RNO-G), but is applicable more general to radio detectors of similar design such as the one proposed for IceCube-Gen2 \cite{Aartsen:2020fgd}. 

Starting in Summer 2021, RNO-G will be the first discovery-scale radio detector array for astrophysical neutrinos. By detecting radio signals from particle showers in the ice sheet of Greenland at distances up to several kilometers, RNO-G will have an effective volume of roughly \SI{\sim 100}{km^3} at the highest energies. To measure the neutrino spectrum or to track energy dependent new physics phenomena, it is essential to be able to reconstruct the energy of the particle showers producing the radio signal, which are used to estimate the neutrino energy.

We have presented a method to reconstruct the shower energy using the data available if a radio signal was detected in only one RNO-G station, incorporating all data from channels deployed deep in the ice. Using this method and MC simulations, a resolution on the shower energy of \SI{\sim 30}{\%} is estimated for hadronic showers after moderate quality cuts.  We have furthermore estimated the resolution on the neutrino energy this would provide, and found this to be about a factor of two on the neutrino energy for a reconstructed shower energy of \SI{1e17}{eV}. The LPM effect causes the uncertainty to increase at higher energies.

For RNO-G an energy-dependent fraction of the detected neutrinos (50\% at \SI{1}{EeV}) are expected to have a signal in more than one station. These events will provide an excellent verification of the method and its systematic uncertainties and can improve the estimation of the shower energy even further.  

The presented method is in general applicable to any type of radio neutrino station in ice that features deep antennas sensitive to two polarizations, distributed at different depths along a string.

\begin{acknowledgements}
We are thankful to the staff at Summit Station for supporting our deployment work in every way possible. Also to our colleagues from the British Antarctic Survey for getting excited about building and operating the BigRAID drill for our project.

We would like to acknowledge our home institutions and funding agencies for supporting the RNO-G work; in particular the Belgian Funds for Scientific Research (FRS-FNRS and FWO) and the FWO programme for International Research Infrastructure (IRI), the National Science Foundation through the NSF Award ID 2118315 and the IceCube EPSCoR Initiative (Award ID 2019597), the German research foundation (DFG, Grant NE 2031/2-1), the Helmholtz Association (Initiative and Networking Fund, W2/W3 Program), the University of Chicago Research Computing Center, and the European Research Council under the European Unions Horizon 2020 research and innovation programme (grant agreement No 805486).
\end{acknowledgements}

\bibliographystyle{JHEP}       
\bibliography{references}   

\end{document}